\documentclass{WileyMSP-template}
\usepackage{amsmath}
\usepackage[utf8]{inputenc}
\usepackage[dvipsnames]{xcolor}
\usepackage[english]{babel}
\usepackage{url}
\usepackage[most]{tcolorbox}
\usepackage{enumitem}
\usepackage{parskip}
\usepackage{hyperref}
\usepackage{chemformula}
\usepackage[version=4]{mhchem}
\usepackage{siunitx}
\usepackage{changepage}
 \usepackage{booktabs}
\DeclareSIUnit{\molar}{M}
\DeclareSIUnit\angstrom{\text {Å}}

\begin{document}

\pagestyle{fancy}

\title{Energetically Driven Structure Matching for Autonomous Total X-ray Scattering Experiments}
\maketitle
\thispagestyle{empty}

\author{Emil J. P. Frost$^{1,2}$}
\author{Martin A. Karlsen$^{1,2}$}
\author{Jonas H. Jensen$^{2,5}$}
\author{Rodrigo Moreno$^{5}$}
\\
\author{Jørn Lambertsen$^{5}$}
\author{Jonathan Quinson$^{6,7}$}
\author{Andy S. Anker$^{1,2,3}$}
\author{Alexander Bagger$^{2,4}$}
\author{Tejs Vegge$^{1,2,\!}$*}

{
\normalsize
\begin{enumerate}[topsep=0pt,parsep=0pt,label={\arabic*}]
    \item Department of Energy Conversion and Storage, Technical University of Denmark (DTU), 2800 Kgs. Lyng\-by, Denmark
    \item Pioneer Center for Accelerating P2X Materials Discovery (CAPeX), Technical University of Denmark \\ (DTU), 2800 Kgs. Lyng\-by, Denmark
    \item Department of Applied Mathematics and Computer Science, Technical University of Denmark (DTU), 2800 Kgs. Lyng\-by, Denmark
    \item Department of Physics, Technical University of Denmark (DTU), 2800 Kgs. Lyng\-by, Denmark
    \item Department of Computer Science, IT University of Copenhagen, 2300 Copenhagen S, Denmark
    \item CICA-Centro Interdisciplinar de Química e Bioloxía, Facultade de Ciencias, Universidade da Coruña, Campus de Elviña, 15008, A Coruña, Spain
    \item Biological and Chemical Engineering Department, Aarhus University, 40 Åbogade, 8200 Aarhus, Denmark
    \item[*] Corresponding email: teve@dtu.dk
\end{enumerate}
}


\begin{abstract}
The emergence of autonomous laboratories motivates rapid conversion of experimental data into reliable atomistic models on time-scales compatible with closed-loop optimization. Here we develop an energetically driven structure matching framework for analysis during ongoing total X-ray scattering experiments. Using data from gold nanoparticles, we match against idealized spherical, octahedral, decahedral, and icosahedral geometries, their machine-learned interatomic potential (MLIP)-relaxed structures, and molecular dynamics (MD) ensembles. Idealized models are fast to generate but can misassign morphology and systematically underestimate size by neglecting surface relaxation, strain, and thermal disorder. MLIP relaxation markedly improves both, while MD ensemble averaging agrees best with experiment. We therefore introduce a hierarchical workflow combining rapid idealized screening with targeted MLIP and MD refinement of top candidates, delivering improved structural feedback without interrupting autonomous operation. This framework provides a route towards autonomous campaigns in which the target structure itself can be updated in response to the evolving energy landscape of structures compatible with the experimental data.
\end{abstract}


\section{Introduction}
The emergence of Self-Driving Laboratories (SDLs)~\cite{canty_science_2025} and Materials Acceleration Platforms (MAPs)~\cite{stier_materials_2024} is transforming materials research by integrating AI and machine learning, automated experimentation, advanced characterization, and high-performance computing into closed-loop discovery workflows~\cite{vogler_autonomous_2024}. SDLs and MAPs rely on advanced workflow orchestrators like PerQueue~\cite{sjolin_perqueue_2024} to accelerate the synthesis, characterization, and optimization of materials by autonomously navigating chemical and structural design spaces while continuously incorporating experimental feedback into autonomous decision-making. More broadly, they represent a shift from human-driven research toward integrated systems capable of conducting materials research at unprecedented speed and scale.

Despite substantial progress in AI-enabled materials discovery, a growing debate has highlighted a fundamental challenge: the distinction between computational prediction or generative design and experimental realization. Even high-profile demonstrations, including DeepMind's GNoME framework~\cite{merchant_scaling_2023} and Berkeley Lab's A-Lab autonomous synthesis platform~\cite{szymanski_autonomous_2023}, generated considerable excitement by proposing and synthesizing large numbers of purportedly novel materials. However, subsequent analyses questioned whether several of the reported compounds were genuinely new, raising broader concerns about structural verification, characterization fidelity, and the criteria by which AI-predicted discoveries should be experimentally validated~\cite{leeman_challenges_2024}. These discussions have reinforced the view that autonomous discovery requires not only prediction and synthesis capabilities, but also robust methods for determining whether experimentally realized structures genuinely correspond to their intended atomic arrangements.

Accurate characterization, therefore, becomes central to autonomous materials discovery. Among the available techniques, total scattering and pair distribution function analysis are particularly attractive because of their sensitivity to atomic structure~\cite{karlsen_importance_2026}. Consequently, scattering measurements are increasingly being considered as key feedback mechanisms for next-generation SDLs and MAPs, providing structural information that can guide autonomous decisions.

We recently demonstrated the use of total X-ray scattering data and pair distribution function analysis as direct feedback for autonomous synthesis through ScatterLab, an autonomous synthesis platform combining robotic experimentation, synchrotron scattering measurements, and Bayesian optimization to synthesize gold nanoparticles (NPs) with user-defined scattering patterns~\cite{anker_autonomous_2026}. By matching experimental scattering patterns with simulated targets in real time, ScatterLab autonomously identified synthesis conditions that produced NPs consistent with targeted decahedral and face-centered-cubic structures. Importantly, the optimization operated directly in scattering space and did not require structural analysis of the experimental data during the loop. This enabled rapid structure-directed synthesis, but also left open a more fundamental question: how uniquely does a given scattering fingerprint map onto an atomic structure? 

Distinct atomic configurations can produce similar and, within experimental resolution, potentially indistinguishable, scattering signatures~\cite{pozdnyakov_incompleteness_2020,maffettone_when_2025}.
Consequently, approaching a simulated target pattern does not necessarily establish that the specific atomic structure used to generate that target has been realized experimentally.
Other studies have addressed this using energy-restrained scattering data analysis~\cite{zarrouk_molecular_2025,anker_autonomous_2025}.
A related question concerns the physical realism of the target itself. Idealized NP models may provide well-defined scattering fingerprints while neglecting surface relaxation, strain, and thermal motion. However, little is known about how the fidelity of atomistic NP models affects their use as targets in autonomous scattering experiments.

\begin{figure}
    \centering
    \includegraphics[width=0.7\linewidth]{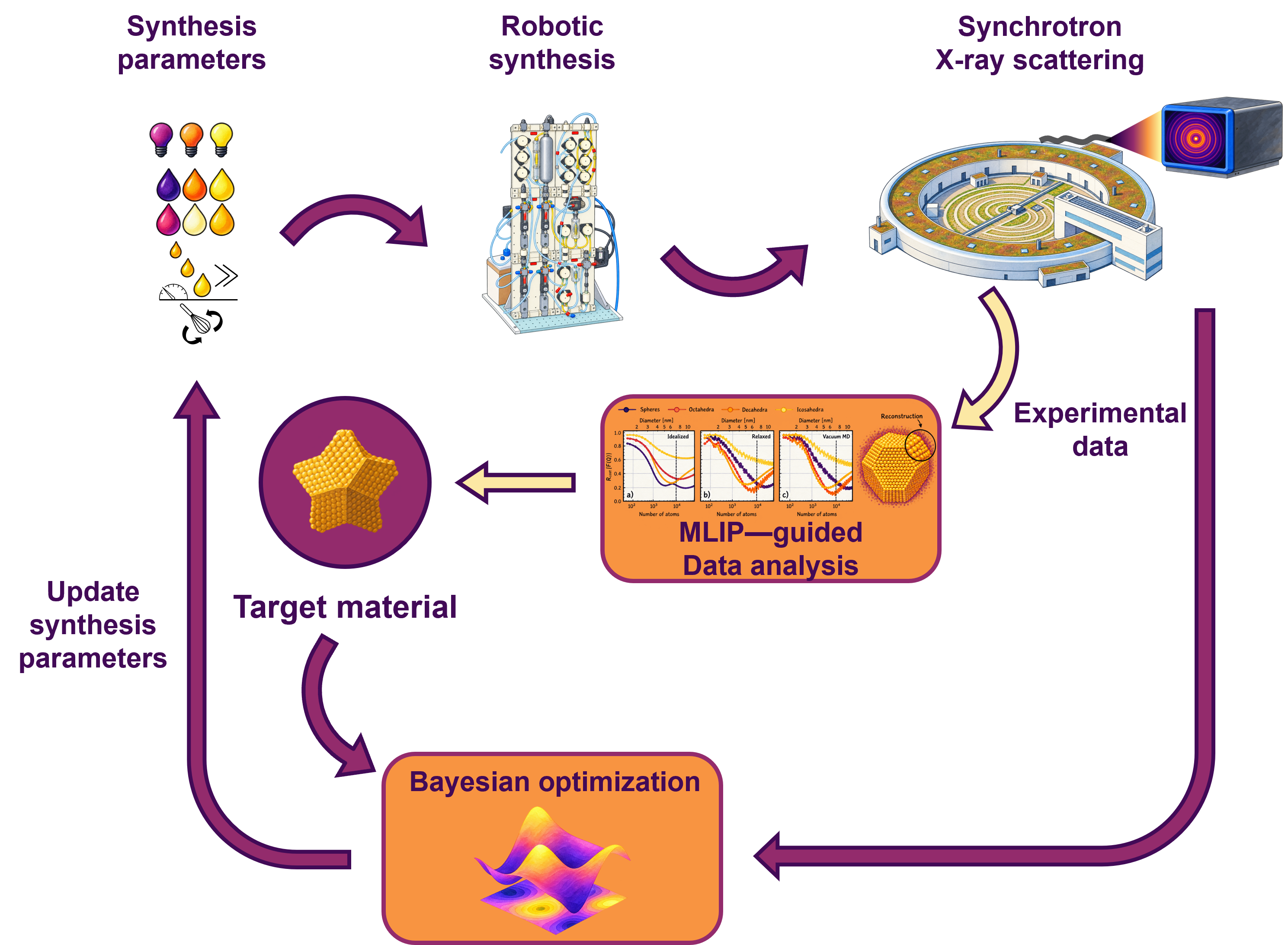}
    \caption{Conceptual extension of the ScatterLab workflow with MLIP-guided in-line structure analysis. Synthesis is performed by the robotic platform, followed by characterization using synchrotron X-ray scattering. Experimental scattering data feed directly into the Bayesian optimizer, which selects synthesis parameters for the next iteration. The yellow arrows indicate the proposed extension to the original ScatterLab workflow: experimental data are analysed against energetically informed candidate structures, and the resulting structural information is fed back to enable updating of the target structure during an autonomous campaign.}
    \label{fig:overview}
\end{figure}

Performing energetically driven scattering data analysis in parallel with the experimental loop could address this limitation. Rather than treating a simulated scattering target as fixed throughout an optimization campaign, experimental data could continuously be compared with candidate structures evaluated on an underlying potential-energy surface (PES). Such an approach could provide an evolving picture of the structural landscape compatible with the experiment, including competing morphologies, particle sizes, and their energetic plausibility. This information could then be used to change the target structure as the experiment progresses, or to extend the optimization objective beyond scattering agreement to structural attributes such as morphology, size, strain, and surface structure. Figure~\ref{fig:overview} illustrates this concept as an extension of the ScatterLab workflow, in which energetically driven scattering data analysis runs alongside synthesis, characterization, and Bayesian optimization and can provide feedback for updating the target structure during an autonomous campaign.

For autonomous experimentation, however, such analysis must occur on the same timescale as synthesis and characterization (i.e. within minutes). It is therefore a challenge that atomistically realistic models generated on a density functional theory (DFT) basis are more computationally expensive in compute time than idealized structural approximations, which makes the direct use of DFT based models unfeasible. Recent advances in machine-learned interatomic potentials (MLIPs) change that by enabling rapid relaxation and molecular dynamics (MD) simulations on timescales matching experiments. This creates an opportunity to move beyond static scattering-pattern matching towards energetically driven structure mining, in which candidate structures are progressively evaluated on the PES to continuously inform autonomous experimental decisions.

Here, we investigate how the fidelity of atomistic NP models influences the interpretation of total X-ray scattering data and develop a computational workflow for energetically driven structure matching on experimentally relevant timescales. We consider candidate gold NPs with spherical, octahedral, decahedral, and icosahedral morphologies using three levels of structural description: idealized structures constructed directly from bulk crystallographic models, structures relaxed on an MLIP-derived PES, and thermally sampled structures obtained through MLIP-driven MD. By systematically comparing these approaches against experimental total X-ray scattering data, we assess how increasing structural fidelity affects morphology assignment, particle-size estimation, and agreement with experiment.

We show that idealized models can misassign NP morphology and systematically underestimate particle size, whereas MLIP relaxation substantially improves both morphology assignment and size estimation, and MD sampling provides the closest agreement with experiment. Based on these findings, we develop a hierarchical structure-mining workflow in which rapid idealized screening narrows the candidate space before targeted MLIP relaxation and MD sampling of the most promising structures. The workflow is designed to operate within experimental cycle times, allowing atomistic structure analysis to proceed in parallel with autonomous synthesis and characterization. By mapping measured scattering patterns to local minima on the PES, this framework provides a route towards autonomous experiments in which structural information can inform both the next synthesis conditions and the target structure itself. By strengthening the connection between experimental observations and atomistic structure, the framework contributes to a broader goal shared across SDLs and MAPs: transforming AI-generated hypotheses into experimentally validated materials discoveries.

\section{Results and Discussion}
For our analysis, we use reference experimental data from an autonomous campaign with the ScatterLab SDL~\cite{anker_autonomous_2026} that generated total X-ray scattering data from gold nanoparticles (see Section~\ref{section:experimental} for experimental details).
We compare the experimental scattering data to a set of simulated spectra from candidate NPs with local face-centered cubic (fcc) crystal structure. The candidate structures are grouped into four morphologies: cleaved single crystals (octahedra), multiply twinned decahedra and icosahedra, as well as spherically terminated single crystals. The first three are common (modified) Wulff constructions~\cite{wulff_xxv_1901,marks_modified_1983} while the last is mostly a convenient construction for comparison.
The Wulff construction is based on the facet-dependent surface energy and minimizes the energy in the thermodynamic limit of large particles. It provides a framework for sampling low-energy candidates from an easily accessible quantity, either directly from DFT calculations or via databases such as the Materials Project~\cite{horton_accelerated_2025, jain_commentary_2013, tran_surface_2016}. 

Once constructed, the atomic positions of each NP are used to simulate X-ray scattering via the Debye equation~\cite{debye_zerstreuung_1915} using DebyeCalculator~\cite{johansen_gpu-accelerated_2024}. Since the NPs are constructed from bulk structures, they are born with a constant nearest neighbour distance and zero strain. These idealized NPs do not necessarily represent the synthesized NP well. This discrepancy is particularly important for multiply twinned NPs, where strain fields and local distortions contribute directly to the total scattering. To accommodate a more faithful description, we employ a DFT-trained MLIP to either perform geometry relaxation or MD of the idealized NPs. Hence, we present three sampling approaches of increasing computational complexity, namely \textit{idealized}, \textit{relaxed}, and \textit{MD sampling}. We refer to the first two sampling methods as \textit{static} and the latter, \textit{dynamic}. 
Geometry relaxation ensures that all candidate structures correspond to local minima on the PES, while MD sampling explores thermally accessible configurations surrounding those minima.
Thermal motion is included in the static methods with a Debye-Waller factor in post-processing.

We find that explicit water solvation during MD has a limited effect on scattering data. The water does not react with the NP, but merely exerts a pressure on it that dampens the motion. This makes explicitly solvated MD at room temperature, $T=300 \, \mathrm{K}$, replaceable by MD in vacuum at a reduced temperature, $T=250 \, \mathrm{K}$ (see SI and discussion in Section \ref{section:misclassification}). This approximation makes MD sampling much more scalable.

\subsection{Fitting Experimental Data}

\begin{figure}
    \centering
    \includegraphics[width=\linewidth]{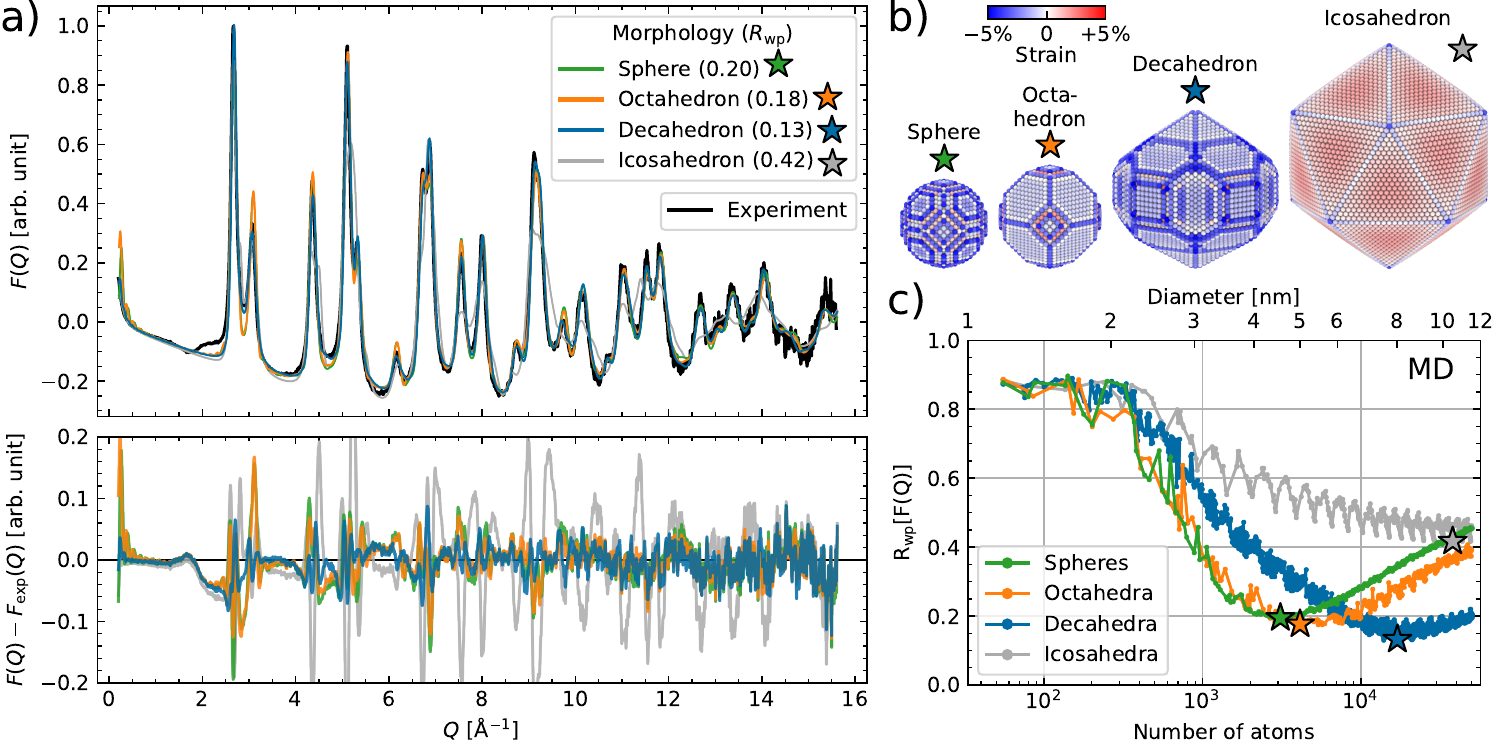}
    \caption{
        Fitting experimental scattering with MD-sampled gold NPs. 
        Panel (a) shows the reduced total scattering function from the experiment (black) 
        along with the best-fitting NP in each considered morphology. The weighted profile residuals, $R_\mathrm{wp}[F(Q)]$, (Equation \eqref{eq:rwp}) between simulated and experimental $F(Q)$ are found in the legend. The $F(Q)$ residuals are shown in the bottom of panel (a). These NPs are the marked minima in (c) whose relaxed geometry (colored by local strain) is shown in (b) in the correct relative sizes. Panel (c) shows goodness-of-fit scans within each morphology as a function of NP size.
    }
    \label{fig:fq-size-scan}
\end{figure}

As an example, Figure~\ref{fig:fq-size-scan} compares all the considered NP models to a representative experimental scattering profile generated using the ScatterLab SDL~\cite{anker_autonomous_2026}.
Figure~\ref{fig:fq-size-scan}a shows the reduced total scattering structure function from the best-fitting NP within each morphology against the experimental spectrum.
The term ``best-fitting'' refers to our goodness-of-fit measure, the weighted profile residual $R_\mathrm{wp}$ (Equation \eqref{eq:rwp}), that achieves its lowest value. 
The decahedral morphology provides the best match to the experiment, as evident from both the $R_\mathrm{wp}$ values and the visual agreement, especially in the low-\textit{Q} region and around the first two peaks.
Scattering from spherical and octahedral NPs show a worse fit but look similar to each other since they are both single-crystal structures.
The icosahedral geometry shows notably poor agreement.
The shoulder on the low-\textit{Q} side of the first peak is missing from all simulated spectra; we attribute this to remaining precursor in the experimental sample.

We show the relaxed geometries of the best-fitting NPs in Figure~\ref{fig:fq-size-scan}b. These structures differ substantially in both size and strain profile (indicated by color). These properties influence both their scattering signatures and catalytic abilities~\cite{sedano_varo_gold_2024}. As such, accurately identifying size and morphology is crucial for understanding the synthesized NPs. Cut-trough views of internal strain are shown in Figure~\ref{fig:strain_sphere_octa} and \ref{fig:strain_deca_ico}.

We estimate the size resolution from size scans across the NPs of each morphology, visualized in Figure~\ref{fig:fq-size-scan}c. The goodness-of-fit curves show well defined minima, allowing for size and uncertainty estimation within each morphology from the minimum position and curvature. Even though the octahedral and decahedral minima appear close in goodness-of-fit value, the $F(Q)$ residuals in Figure~\ref{fig:fq-size-scan}a indicate that the two morphologies can be distinguished.
As a pragmatic criterion for estimating the size uncertainty, we assume a tolerance of $\Delta R_\mathrm{wp}=0.01$ around each minimum. Applying this criterion, the optimal octahedral and decahedral diameters are $d=\SI{5.3 \pm 0.5}{nm}$ and $d=\SI{8.2 \pm 0.4}{nm}$, respectively. These values are estimated from the mean and standard deviation of the NP sizes lying within $\Delta R_\mathrm{wp}=0.01$ of each minimum (22 octahedral and 52 decahedral candidate NPs fulfill this criterion). These size disparities highlight the importance of considering morphology and size jointly when determining the underlying atomic structure, as different morphologies can exhibit comparable coherent structural length scales despite differences in overall particle size.

Additionally, we observe discontinuous behavior in the loss curves of Figure~\ref{fig:fq-size-scan}c. This arises from different surface terminations that exhibit distinct strain responses during PES sampling. These variations in strain directly influence the overall scattering signal. While X-ray scattering probes the particle as a whole (bulk-sensitive), surface alterations do produce measurable scattering differences. These effects are particularly pronounced for \textit{small} particles, where surface atoms contribute a larger fraction to the combined signal.

\subsection{Comparing Sampling Methods}

\begin{figure}
    \includegraphics[width=\linewidth]{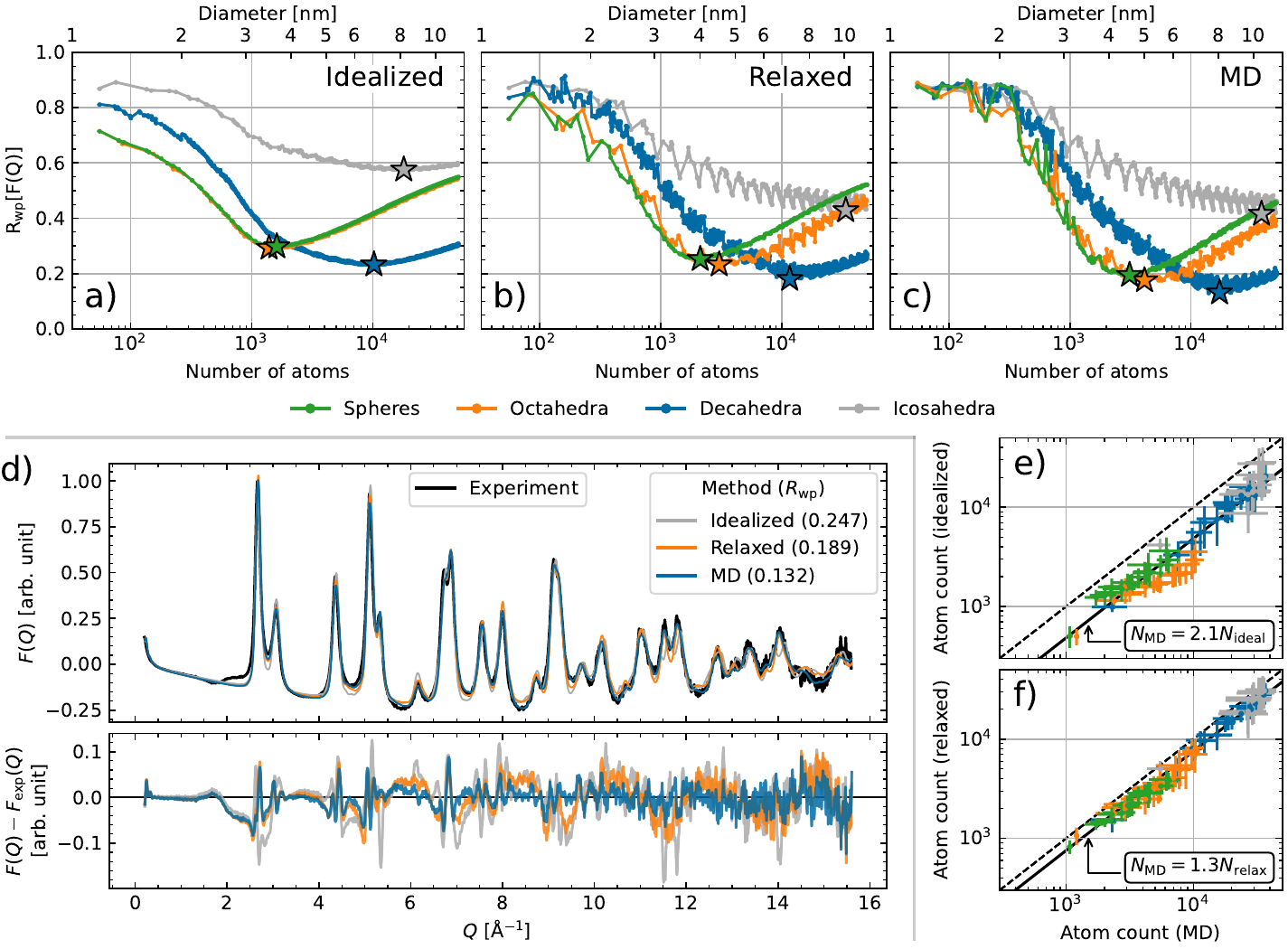}
    \caption{Comparison between sampling approaches. (a-c) Size scans with (a) idealized, (b) relaxed, and (c) MD-sampled NPs (same as Figure~\ref{fig:fq-size-scan}c). (d) Reduced total scattering structure functions between the methods using the best-fitting NP in the MD approach (blue star in (c)). Residuals from experiment are shown in the bottom panel. (e-f) Size predictions of 20 experimental samples shown as parity plots between methods. Error bars show the mean and standard deviation of the distribution of particle sizes within $\Delta R_\mathrm{wp}=0.01$ of the minimum. The dashed line is the parity line and the full line is the best linear fit (slope written in inset).}
    \label{fig:method-comparison}
\end{figure}

\begin{figure}
    \centering
    \includegraphics[width=0.55\linewidth]{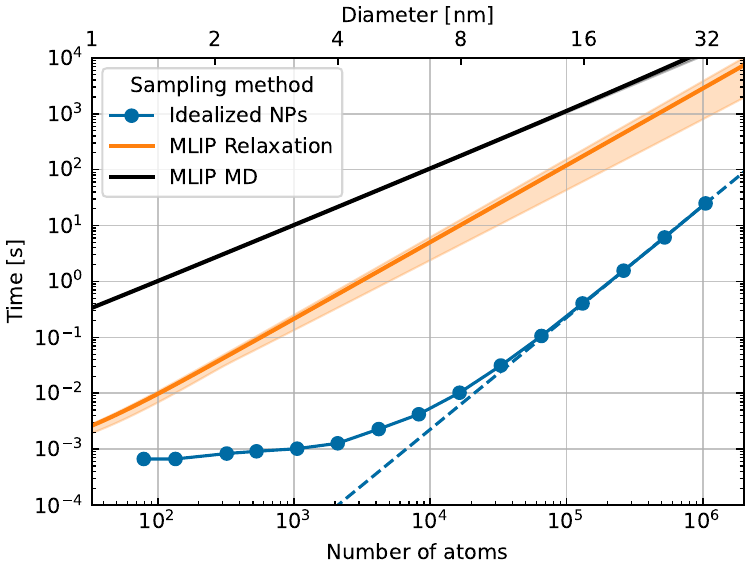}
    \caption{Time usage for simulating scattering from a NP on an NVIDIA H200 for the three proposed sampling methods: idealized, geometry-relaxed, or MD-sampled NPs. Idealized sampling only computes simulated scattering -- the others also update geometries with a finetuned MACE-MATPES-PBE-0 model as the MLIP.
    We assume an MLIP troughput of \SI{400 000}{atoms \times steps / s} (see Figure~\ref{fig:md_benchmark}c) and use 4000 energy/force evaluations for MD (\SI{10}{ps} equilibration + \SI{10}{ps} sampling with a \SI{5}{fs} time step).
    Shaded areas indicate uncertainty in the number of required relaxation steps; the boundary is from power law fits to the number of relaxation steps required for each morphology (using the data in Figure~\ref{fig:relax_steps_size}).
    The dashed line is a quadratic fit to the last five points for idealized sampling of spherical NPs.
    All methods make use of the binned Debye equation (Equation \eqref{eq:binned_debye_single}) with a grid of $N_Q=2067$ \textit{Q}-values, as in the experiments, and a bin width of \SI{0.001}{\angstrom}.}
    \label{fig:methods_time}
\end{figure}

We evaluate each sampling method by three measures. First, we quantify the goodness-of-fit to the experimental data. This gives an indication of how realistic the simulated data is. Second, we use parity plots to determine if all approaches yield consistent size estimates and size uncertainties. Third, we estimate the required computational resources.

The three approaches are compared in Figure~\ref{fig:method-comparison}. Figure~\ref{fig:method-comparison}(a-c) show the size analysis of the three sampling methods. Three observations are apparent as we move from idealized to relaxed to MD-sampled NPs.
First, the minima are consistently lowered within all morphologies. As computational cost increases, the simulated data better match the experimental.
Second, idealized sampling shows almost no curve broadening, meaning surface changes are not resolvable with simulated scattering from idealized NPs. Differences only appear when structures are mapped to minima on the PES. In particular, idealized sampling can not differentiate the two types of single-domain morphologies, octahedra and spheres, as these only differ in surface termination. The MLIP-based methods cleanly separates the two.
Third, the minima shift to larger sizes. In fact, across 20 experimental datasets generated with the ScatterLab SDL~\cite{anker_autonomous_2026}, the idealized approach predicts only about half as many atoms in the synthesized particle compared to MD sampling (see Figure~\ref{fig:method-comparison}e).
Due to the constant neighbor distance of idealized NPs that extends to the particle edge, NPs need to be \textit{smaller} than the equivalent MD-sampled NPs to match coherent domain sizes.
Relaxed sampling estimates about 25\% fewer atoms than MD sampling and overall yield more consistent and predictable size estimates than idealized sampling (see Figure~\ref{fig:method-comparison}e-f).
Since both MLIP-based methods sample the same minima on the PES, good correlation is also expected between relaxed and MD sampling.

Figure~\ref{fig:method-comparison}d compares the simulated reduced total scattering structure functions across the three approaches using the best-fitting MD-sampled NP. To this comparison, we make two remarks. First, the idealized structure results in sharper features than the more disordered relaxed and MD-sampled structures with modified strain. Second, the static methods fail to capture the decay of $F(Q)$ at high $Q$; the MD-sampled structure exhibits a more realistic decay and a flatter residual than the two static methods. Thus, temperature effects are not properly captured with a Debye-Waller correction.
Direct sampling of the PES through MD is more accurate, although computationally more expensive.

The time usage for the three sampling methods is summarized in Figure~\ref{fig:methods_time} in the limit of negligible MLIP overhead, i.e. perfect linear weak scaling. Relaxed sampling is 1--2 orders of magnitude faster than MD, with another 2--3 orders of magnitude to idealized sampling. Note that relaxed sampling does not quite scale linearly with the number of atoms since more relaxation steps are required for larger particles (see Figure~\ref{fig:relax_steps_size}).
Even though the Debye equation scales quadratically, the time required to sample the geometries with MLIPs far outweighs the time used to compute the simulated scattering data. Thus, the time-limiting factor in this study is MLIP evaluation.

\subsection{Misclassification from Idealized Sampling}
\label{section:misclassification}

\begin{figure}
    \centering
    \includegraphics[width=0.75\linewidth]{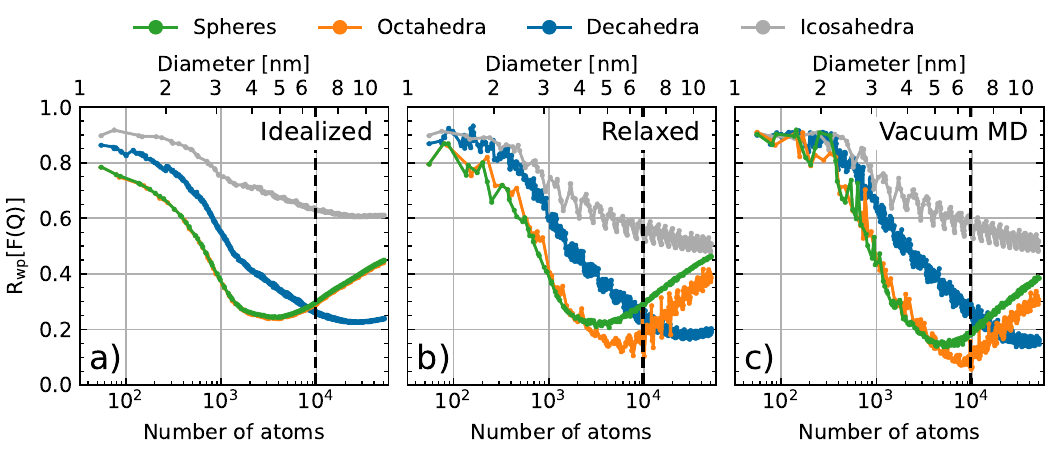}
    \hfill
    \raisebox{0.15\height}[0pt][0pt]{\includegraphics[width=0.24\linewidth]{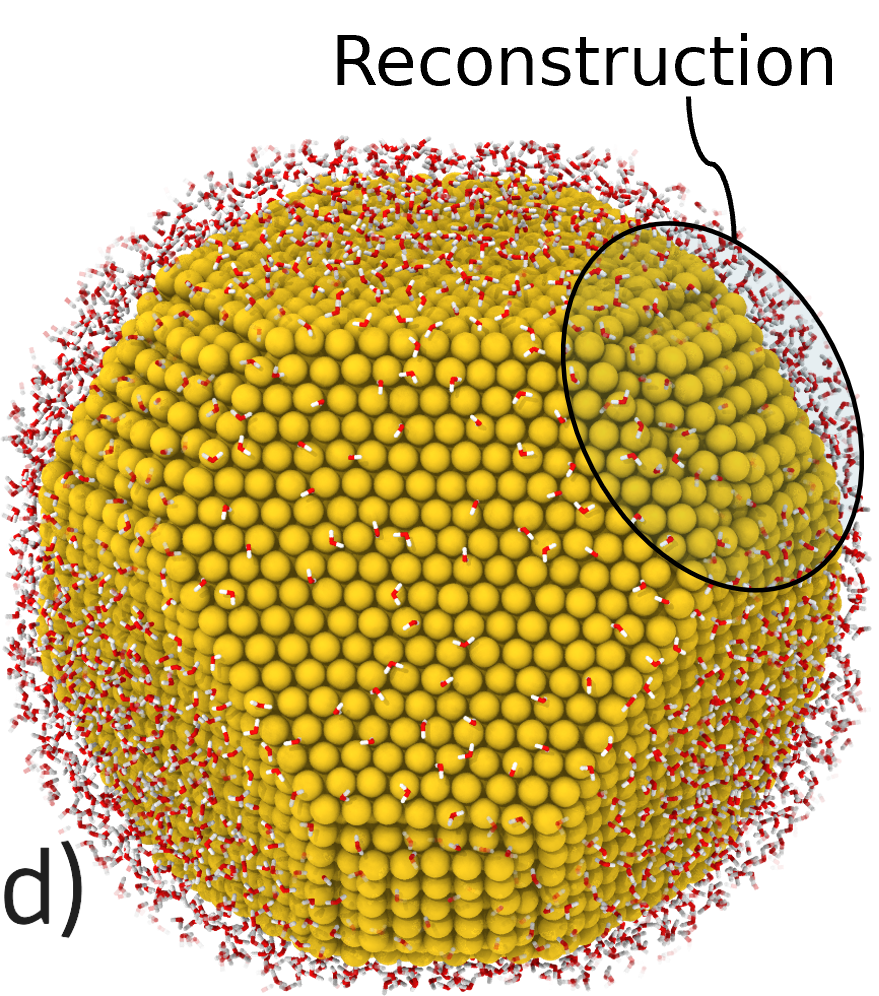}}
    \caption{Structure matching for a 7 nm truncated octahedron sampled by MLIP-driven MD in explicit water. Panels (a-c) show goodness-of-fit size scans computed from (a) idealized, (b) relaxed, and (c) MD-sampled (at $T=\SI{250}{K}$) configurations in vacuum; the dashed vertical line marks the simulated particle size. (d) Snapshot of the solvated reference system (water partly omitted for clarity). Reference scattering was averaged over the final 10 ps of a 100 ps MD NVT trajectory at $T=\SI{300}{K}$ and $\SI{1.0}{g/cm^3}$ water density.}
    \label{fig:solvated_octa}
\end{figure}

Idealized sampling is computationally inexpensive and useful for rapid screening, but produces misleading size estimates and, in some cases, incorrect morphology assignments. Relaxed and MD-sampled structures, which incorporate surface reconstruction and strain effects, systematically improve goodness-of-fit and predict larger particle sizes than idealized sampling. To be able to tell the difference between decahedral and octahedral morphologies in experimental samples, it is essential that the analysis method is able to correctly and confidently separate the two classes. Decahedra are multiply twinned particles with five fcc domains of different orientation. Each domain produces scattering that is reminiscent of smaller, single domain NPs. Through the lens of X-ray scattering, large decahedra resemble smaller octahedra with matching domain size. This is the reason the two morphologies may be confused in analyses.

Figure~\ref{fig:solvated_octa} presents a concrete example of morphology misclassification using simulated scattering data from a 7 nm octahedron in explicit water, sampled via MD at $T=\SI{300}{K}$ with our finetuned MACE model~\cite{batatia_mace_2022, batatia_design_2022, kaplan_foundational_2025, batatia_foundation_2025}. Two key observations emerge from this comparison of sampling methods. First, consistent with the general trend, goodness-of-fit improves and estimated sizes increase from idealized to relaxed to MD sampling. The best fitting relaxed NP is located at the correct size, though the overall curve appears slightly shifted towards smaller sizes. Second, the idealized sampling method shows poor differentiation between octahedral and decahedral morphologies, slightly favoring decahedra. In contrast, relaxed and MD sampling methods confidently identify the correct octahedral morphology. We also note that vacuum MD correlates well with the solvated MD reference, indicating that for gold NPs, the effect of water is well approximated by a reduction in temperature of the vacuum-exposed MD from $T=\SI{300}{K}$ to $T=\SI{250}{K}$. For more reactive systems, however, explicit solvent treatment may be essential to accurately capture the interactions and dynamics of the NPs.

Figure~\ref{fig:solvated_octa} represents an experimentally relevant case where idealized sampling fails. The simulated octahedron is of similar size to the 8 nm decahedron fitted to the experimental data in Figure~\ref{fig:fq-size-scan} and is predicted to have lower potential energy in vacuum than the decahedral candidates at this size (see Figure~\ref{fig:energy_vs_size}). Due to the inert nature of gold, the vacuum Wulff construction is expected to be a good approximation of the thermodynamically stable solvated NP. Hence, it represents our best estimate of a thermodynamically stable 7 nm octahedral candidate.
Additionally, it is kinetically stable with only small changes to the particle edges during the explicitly solvated MD.
The close-packed (111)-facets are unaltered in MD but the truncated corners may change slightly. In Figure~\ref{fig:solvated_octa}d, the upper right corner shows signs of small surface reconstructions when compared to the other two corners. In general, open facets such as (100) are not kinetically stable during our MD simulations. Both in vacuum and in water, atoms are pushed out at (100)-terminations, transforming them into more rounded, closer packed terminations. Compared with idealized and relaxed structures, MD-sampled structures have more rounded edges and corners.

Comparisons with synthetic data from vacuum MD trajectories reveal many cases where idealized sampling predicts a different morphology than relaxed and MD sampling. For the idealized approach, small decahedra are often misidentified as icosahedra, while small octahedra and large icosahedra tend to be misclassified as decahedra, in some cases with high apparent confidence (see Figure~\ref{fig:confusion_matrix} and \ref{fig:family_confidence}). In contrast, the relaxed and MD-sampled methods predict consistent morphologies. This is not surprising since MD sampling is initialized from the relaxed geometry. Apart from small surface reconstructions for kinetically unstable candidates, the added effect of MD sampling is a more faithful description of vibrational modes.

\subsection{A Funnel Approach to Efficient Structure Mining}
As shown above, MLIP-based methods that sample the PES provide better agreement with experiment, but at a higher computational cost. The less accurate idealized sampling, in contrast, has the great advantage that it requires no PES evaluation. Thus, with an efficient implementation of the Debye equation, quick comparisons can be made with experiments. For simple, small, single-element NPs, an exhaustive search among candidates is tractable with MD sampling in a reasonable amount of time. However, for more complicated systems, where the search space is impractically large, a more efficient time usage is needed. Therefore, we propose to arrange the three sampling methods in a funnel approach to leverage the strengths of each. Idealized sampling is followed by relaxed sampling to progressively narrow the set of promising candidates, allowing the available computational time to be focused on MD sampling the most relevant structures. The protocol is intended for online structure matching with only short time for analysis.

We envision autonomous experiments and analyses running in parallel, offset by one sample so that each analysis is completed within a single experimental cycle. The resulting mapping from synthesis conditions to structure can guide subsequent experiments and enable multi-objective optimization of particle \textit{features} such as morphology, size, and surface structure, instead of scattering alone. The target material could also be changed on the fly, based on more realistic, synthesizable NPs (as depicted in Figure~\ref{fig:overview}). For time-constrained online analysis, we outline this funnel approach below.

\begin{tcolorbox}[
    colback=blue!5,
    colframe=blue!45!black,
    boxrule=0.6pt,
    arc=3mm,
    left=6mm,
    right=6mm,
    top=4mm,
    bottom=4mm
]
\begin{center}
    \textbf{Funnel Approach}
\end{center}

\begin{enumerate}[leftmargin=*, itemsep=0.5em]
    \item \textbf{Idealized sampling.}
    Quick size scans locate the approximate optimum for each morphology and establish an upper size limit for subsequent MLIP calculations. A loss threshold, $\Delta R_\mathrm{wp}^\mathrm{idealized}$, above the minimum loss defines the maximum number of atoms, $N_\mathrm{max}$, in Step 2.

    \item \textbf{Relaxed sampling.}
    A subset of candidates within the reduced size range is relaxed. Poorly matching morphologies are discarded and the sampling window is refined. 

    We fit the loss curve with the empirically motivated function
    \begin{equation}
        \label{eq:rwp_fit}
        \hat{R}_\mathrm{wp}(t) = c_0 + c_1t + c_2t^2 + c_3t^3, \quad t = \log N.
    \end{equation}
    We define the refined window as $\Omega = \{N \mid \hat{R}_\mathrm{wp}(t(N))<\hat{R}_\mathrm{wp}(t_\mathrm{min})+\Delta R_\mathrm{wp}^\mathrm{relaxed}\}$, where $t_\mathrm{min}$ is the fitted minimum and $\Delta R_\mathrm{wp}^\mathrm{relaxed}$ is chosen such that $\Omega$ comfortably contains the MD minimum.

    \item \textbf{Guided MD sampling.}
    Candidates with $N\in\Omega$ are first ranked using relaxed sampling. Then the most promising candidates are forwarded to MD sampling for final evaluation.

    We bin candidates in the $\Omega$ window and use the width-normalized deviation from the fitted loss as predictor of MD sampling outcome.
    \begin{equation}
        \label{eq:z}
        z = \frac{R_\mathrm{wp}-\hat{R}_\mathrm{wp}(t(N))}{W(N)},
    \end{equation}
    where $W(N)=\sqrt{\frac{1}{K}\sum_{k=1}^K[R_{\mathrm{wp},b}-\hat{R}_\mathrm{wp}(t(N_b))]^2}$ is the loss curve width in the bin that $N$ belongs to. We relax randomly sampled candidates from $\Omega$ and perform MD sampling of them in ascending order of $z$ to sample the bottom of the loss curve until the time limit is met.
\end{enumerate}
\end{tcolorbox}

\subsubsection{The Funnel in Action}
As a proof of concept, we apply the proposed workflow to experimental scattering data retrospectively.
We assume access to eight NVIDIA H200 GPUs and a 15-minute search window, consistent with the cycle time reported for previous autonomous synchrotron X-ray scattering experiments~\cite{anker_autonomous_2026}.
Further, we assume each geometry relaxation uses 300 MLIP energy/force evaluations and MD uses 4000: \SI{10}{ps} equilibration followed by \SI{10}{ps} production sampling. The relationship between particle size and compute time is summarized in Figure~\ref{fig:methods_time}. Convergence tests for the relaxation and MD protocols are provided in Figure~\ref{fig:relax_steps_hist} through \ref{fig:fq_md_convergence_violin}.

We begin the funnel approach by screening the candidate pool with idealized sampling (Figure~\ref{fig:method-comparison}a). Applying $\Delta R_\mathrm{wp}^\mathrm{idealized}=0.02$ sets the upper size limit to $N_\mathrm{max}=50\,000$. The threshold needs to match the experimental data. It may be set to a large value initially and later refined.

We then scan 16 octahedra, decahedra, and icosahedra distributed across this range using relaxed structures, requiring approximately 20 seconds per morphology (see Figure~\ref{fig:relaxed_size_scan}). Spherical particles are excluded \textit{a priori} because they are not physically relevant. We fit each resulting goodness-of-fit curve with a cubic polynomial in $\log N$ (Equation~\eqref{eq:rwp_fit}) and use the fits to refine the sampling window. 
We retain the morphology with the lowest relaxed loss and discard poorly fitting morphologies -- relaxed and MD sampling identify the same morphology in our synthetic tests (see Figure~\ref{fig:confusion_matrix}b). If the morphology confidence is low, we retain all competitive morphologies. In this case, we confidently pick only the decahedra for Step~3. Using $\Delta R_\mathrm{wp}^\mathrm{relaxed}=0.02$ relative to the fitted minimum gives a refined size window of $\Omega=[8\,000,29\,000]$, corresponding approximately to diameters of 6--\SI{10}{nm}.

We pass $\Omega$ to guided sampling for the remaining 14 minutes. Randomly sampled candidates in the window are relaxed and ranked according to \textit{z} (Equation \eqref{eq:z}). Structures that lie near the minimum of the relaxed-sampling loss curve also tend to perform well in MD sampling. Hence, we balance adequate sampling of the minimum and exploration of candidates by choosing a ratio of ten relaxed samples to one MD sample. This strategy is justified by the strong correlation between relaxed and MD scattering (see Figure~\ref{fig:time_constrained_md}a). In general, the optimal ratio depends on their correlation and relative computational costs.

\begin{figure}
    \centering
    \includegraphics[width=\linewidth]{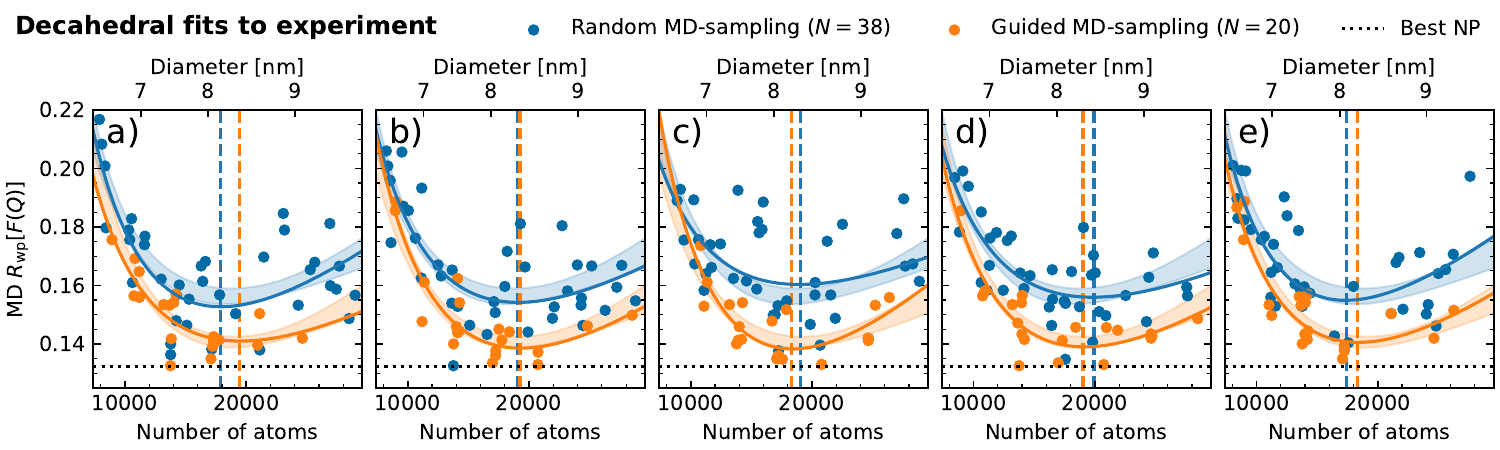}
    \caption{Comparison of random and guided MD-sampling as the final step in a 15 minute time-constrained funnel search for the best fitting NPs to experiment. Panels (a-e) show five separate instances of such searches in a bootstrapping analysis with 10\! 000 trials. The full lines are fits to a quadratic function in log space: $\hat{R}_\mathrm{wp}(t)=c_0+c_1t+c_2t^2, \ t=\log N$. The dashed lines are the fit minima, $N_0$, of the particular trial, while the shaded areas are $1\sigma$ prediction intervals (PIs) on the fit. The $1\sigma$ PIs on the minima are $N_0^\mathrm{random} \in [17\,700, 19\,700]$ and $N_0^\mathrm{guided} \in [18\,200,20\,000]$, or, in terms of diameters, $d_0^\mathrm{random} \in [8.1, 8.4] \ \mathrm{nm}$ and $d_0^\mathrm{guided} \in [8.2, 8.5] \ \mathrm{nm}$. Note that this uncertainty comes from different sub-samplings of candidates; it is \textit{not} a statement about confidence in the true size of the synthesized particle.
    }
    \label{fig:funnel}
\end{figure}

Figure~\ref{fig:funnel} shows an example of five trials of MD sampling in the $\Omega$ window. Guided sampling substantially outperforms random sampling within the search interval. Although random MD sampling uses nearly twice as many points, both approaches are expected to yield similar optimal size estimates. The key difference is the \textit{quality} of the sampled candidates: guided sampling produces more excellent candidates than random sampling.
Figure~\ref{fig:time_constrained_md}b summarizes the expected goodness-of-fit of the sampled candidates from 10\,000 such trials. For this particular experimental dataset, 15 minutes of guided sampling yields 20 MD-sampled NPs of which the best lies within $\Delta R_\mathrm{wp}=0.005$ of the best candidate in the full collection in 95\% of the trials. This difference is small enough that the corresponding scattering functions are visually indistinguishable.
As experiments continue, the growing database of simulated X-ray spectra may further accelerate the analysis.

\section{Conclusion}
Real-time structure matching has the potential to provide structural information during autonomous synthesis experiments, but its reliability depends on the fidelity of the atomistic models used to interpret experimental scattering data. Here, we show that the choice of structural model directly influences both morphology assignment and NP size estimates from total X-ray scattering.
Idealized NPs enable efficient initial screening but systematically yield smaller particle size estimates and may misclassify morphology, because they neglect surface reconstruction and internal strain.
MLIP-based geometry relaxation improves agreement with experiment by mapping idealized particles onto minima on the PES with more realistic strain and surface ordering.
MD sampling provides a further improvement by averaging over thermally accessible configurations on the PES and yields the closest agreement with experiment, albeit at greater computational cost.

To make this level of structural fidelity compatible with autonomous experimentation, we developed a hierarchical funnel workflow that leverages the strengths of all three sampling methods.
Under experimentally relevant computational time constraints, the workflow rapidly narrows the candidate space in morphology and size before directing computational effort towards high-fidelity MD simulations of the most promising structures.

This provides a practical route for atomistic structure analysis to proceed in parallel with synthesis and characterization, rather than relying solely on direct scattering-pattern agreement for autonomous feedback. 
More broadly, connecting experimental scattering fingerprints with energetically informed candidate structures provides additional information about the structural landscape compatible with an experiment. In-loop, energetically driven structure matching, therefore, provides a route towards higher-fidelity structural feedback for self-driving laboratories and materials acceleration platforms.
We envision that a powerful in-loop structure characterization tool in future autonomous synthesis campaigns could enable direct steering towards target \textit{features} such as morphology, size, strain, or surface effects instead of a single target \textit{material}.

\section{Methods}
\subsection{Experimental}
\label{section:experimental}
\subsubsection{Synthesis}
Nanoparticle solutions were synthesized and injected for X-ray acquisition using
the Modular Experimentation Platform (MODEX) in a self-driving laboratory, as 
described by Anker \textit{et al.}~\cite{anker_autonomous_2026}.
A background synthesis was performed under the same conditions, with the gold-precursor volume replaced by water.

All chemicals were used as received. Stock solutions used for the synthesis were prepared from high-purity Milli-Q water (\ce{H2O}, Millipore, resistivity \SI{>18.2}{\mega\ohm\cdot\centi\meter}); chloroauric acid trihydrate (\ce{HAuCl4*3H2O}, 99\%, BLD Pharmatech; \SI{20}{\milli\molar}); lithium citrate tribasic tetrahydrate (\ce{Li3C6H5O7*4H2O}, $\geq \! 99.5 \%$, BioUltra, Sigma-Aldrich; \SI{500}{\milli\molar}); lithium borohydride (\ce{LiBH4}, 95\%, Thermo Scientific Chemicals; \SI{200}{\milli\molar}); lithium hydroxide (\ce{LiOH}, $\geq \! 98.0\% $, ACS reagent, Sigma-Aldrich; \SI{500}{\milli\molar}); and polyvinylpyrrolidone (PVP, Sigma-Aldrich, average molecular weight 10,000; \SI{500}{\milli\molar} on a monomer basis). For PVP, a molar mass of \SI{111.14}{\gram\per\mole}, corresponding to the vinylpyrrolidone repeating unit, was used to calculate the concentration.
\textit{Safety note:} \ce{LiBH4} is highly reactive and flammable and should be handled in a well-ventilated area. When preparing aqueous solutions, \ce{LiBH4} should be added gradually to water rather than adding water to \ce{LiBH4}, as the latter may result in vigorous reactions and ignition. Decomposition of \ce{LiBH4} generates gases; therefore, closed containers may become pressurised. All chemicals and waste should be handled and disposed of in accordance with the applicable local safety regulations and procedures.

\subsubsection{Synchrotron X-ray Scattering}
Synchrotron X-ray scattering experiments were performed at the DanMAX beamline, 
MAX IV, Lund, Sweden.
The X-ray wavelength was \SI{0.3542}{\angstrom} (approximately \SI{35}{\kilo
\electronvolt}).
The X-ray beam dimensions were $\SI{926.4}{\micro\meter}\times\SI{780.0}{\micro
\meter}$ (full-width at half-maximum).
A Dectris Pilatus3 X 2M CdTe area detector was used for data collection (pixel 
size of $\SI{172}{\micro\meter}\times\SI{172}{\micro\meter}$, active area of 
$1475 \times 1679$ pixels).
Fused silica capillaries with an inner diameter of \SI{0.7}{\milli\meter} were 
used for all measurements. Capillaries were held static, orthogonal to the X-ray
beam, and were probed in Debye-Scherrer geometry using a rapid-acquisition pair distribution function
(RA-PDF) approach~\cite{chupas_ra-pdf_2003}.
Calibration of the experimental geometry was performed by measuring Si NIST SRM 
640f powder loaded in a fused silica capillary. For the Si calibrant, an empty 
fused silica capillary was measured as background for subtraction.
Data were acquired for \SI{300}{\second}. The pyFAI software was used to 
calibrate the experimental geometry and create a static mask for detector gaps, 
dead pixels, and beamstop~\cite{kieffer_pyfai-calib_2020}.
To increase the accessible $Q$-range, the area detector was tilted~\cite{burns_inclined-detector_2023}.
The sample-to-detector distance and the detector tilt angle were calibrated to 
\SI{179.788}{\milli\meter} and \SI{44.78}{\degree}, respectively.
The $Q$-range captured by the area detector was from \SI{0.2}{\per\angstrom}
to \SI{22.0}{\per\angstrom}. This was also the $Q$-range used for the Si 
calibrant. The PDFs of the samples were obtained using a $Q$-range from 
\SI{0.2}{\per\angstrom} to \SI{15.6}{\per\angstrom}.
Samples and their corresponding solution-based backgrounds were measured for 
\SI{60}{\second} each.
Azimuthal integration was performed using the azint pipeline available at DanMAX 
and MAX IV~\cite{jensen_azint_2022}.
Both solid-angle correction and polarization correction (factor 0.965) were 
included during the azimuthal integration.
Further data processing, including background-subtraction, normalization, 
\textit{ad hoc} corrections for incoherent scattering, and inverse Fourier 
transformation, was performed using the PDFgetX3 algorithm~\cite{juhas_pdfgetx3_2013}.
To estimate instrumental contributions to the PDFs, the PDF of the Si calibrant 
was modelled using DiffPy-CMI~\cite{juhas_diffpy-cmi_2015}.
Instrumental parameters were estimated to 
$Q_{\mathrm{damp}}=\SI{0.023}{\per\angstrom}$ and $Q_{\mathrm{broad}}=\SI{0}
{\per\angstrom}$.

\subsection{Simulating X-ray Scattering}
To simulate scattering from NPs, we used the Debye equation to compute the scattering intensity \cite{debye_zerstreuung_1915}.
\begin{equation}
    I(Q) = \sum_{ij} f_i(Q) f_j(Q) \frac{\sin(Qr_{ij})}{Qr_{ij}}.
    \label{eq:debye}
\end{equation}
Here, $Q$ is the magnitude of the scattering vector, and the sum runs over all pairs of atoms with form factors $f_i(Q)$ and $f_j(Q)$, separated by distances $r_{ij}$. For a single-element system such as gold NPs, $f_i(Q) = f_j(Q) = f(Q)$ and this reduces to
\begin{equation}
    \label{eq:debye_single}
    I(Q) = f^2(Q) \sum_{ij} \frac{\sin(Qr_{ij})}{Qr_{ij}}.
\end{equation}

Evaluating this expression for $N$ atoms and $N_Q$ $Q$-values has complexity $\mathcal{O}(N_QN^2)$. We accelerated the calculation by binning interatomic distances: for $B$ bins with centers $r_b$ and counts $w_b$, we approximated
\begin{equation}
    \label{eq:binned_debye_single}
    I(Q) \simeq f^2(Q) \sum_{b=1}^{B} w_b \frac{\sin(Qr_{b})}{Qr_{b}},
\end{equation}
which reduced the cost to $\mathcal{O}(N_QB)$. The limiting cost of this approach is the distance binning rather than the summation. Using binary search to assign distances to bins yields a complexity of $\mathcal{O}(N^2\log B)$. Larger particles require more bins, and, for a fixed bin width, $B \propto N^{1/3}$. Benchmarks (see Figure \ref{fig:debye_benchmark}) show that the binning approach is two orders of magnitude faster. GPUs efficiently parallelize the computation, enabling simulations with a million atoms in tens of seconds. A similar approximation is possible for multi-element systems with histograms for all element pairs. We did not avoid the quadratic scaling with the number of atoms, but we reduced the overall prefactor. In the SI, we show how $I(Q)$ computed from Equation \eqref{eq:binned_debye_single} converges to the result of Equation \eqref{eq:debye_single} as the bin width decreases.

Further approximations of the scattering intensity could be valuable for even larger systems. In such cases, one needs to consider fewer than all $\sim \! N^2/2$ interatomic distances. 
One option is the introduction of a distance cutoff for linear scaling~\cite{zarrouk_linear-scaling_2025}.
Another possibility is to approximate the bin weights ($w_b$ in Equation \eqref{eq:binned_debye_single}) by stochastically sampled atom pairs (perhaps only $\sim \! N$), instead of considering all pairs. Such approximations should be viable for smooth distributions of distances. Since the bottleneck in our explored size range was the MLIP, distance cutoffs or implementation attempts of subsampling were not made; we leave this as an idea for future development.

For the static sampling methods, idealized and relaxed sampling, thermal atomic motion was accounted for via the Debye-Waller term for isotropic harmonic vibrations \cite{debye_interferenz_1913, waller_zur_1923}:
\begin{equation}
    I'(Q) = I(Q) \exp\left(-\frac{B_\mathrm{iso}Q^2}{8\pi^2}\right).
    \label{eq:idp}
\end{equation}
Here $B_\mathrm{iso}$ is the isotropic atomic displacement parameter (ADP), which for harmonic, isotropic vibrations is proportional to the mean squared displacement $\langle r^2\rangle$. From comparisons with the dynamic MD sampling method (see Figure~\ref{fig:fitted_biso} and \ref{fig:rwp_biso_scan} in the SI), the ADPs for idealized and relaxed sampling were fixed to $B_\mathrm{iso} = \SI{0.95}{\angstrom^2}$ and $B_\mathrm{iso} = \SI{0.69}{\angstrom^2}$, respectively.
Our finding show that fixating the ADP improves size discrimination and, hence, morphology classification (compare Figure \ref{fig:method-comparison}a and \ref{fig:method-comparison}b with Figure~\ref{fig:rwp_scan_free_biso}). NP size and ADP are correlated since they both affect the peak heights of $F(Q)$. In the absence of synthetic data to compare with, fitting the ADP of each structure directly to experimental data is still viable but less effective.

From the scattering intensity, we computed the reduced total scattering structure function
\begin{equation}
    F'(Q) = \frac{Q}{Nf^2(Q)} I'(Q).
\end{equation}
To account for instrumental resolution, we convolved $F'(Q)$ with a Gaussian kernel, $\kappa(Q; Q_\mathrm{damp})=\exp(-Q^2/2Q_\mathrm{damp}^2)/\sqrt{2\pi Q_\mathrm{damp}^2}$, at the end.
\begin{equation}
    F(Q) = \int \mathrm{d}Q' \, F'(Q-Q') \kappa(Q'; Q_\mathrm{damp}).
\end{equation}
The real-space and reciprocal-space scattering functions contain the same information. In this work, we chose to fit only $F(Q)$.

We introduced two free parameters for each candidate structure when matching simulated and experimental scattering: an overall scaling of the positions, $c_r$, which reflected lattice-constant errors in DFT, and an intensity rescaling of $F(Q)$, $c_F$. In Equation \eqref{eq:binned_debye_single}, distances appear only in products with $Q$. Hence, the summation is left invariant under the transformations $r_{ij} \to c_r r_{ij}$, $Q \to Q/c_r$. This gives an efficient way to fit $c_r$ without recomputing the summation including binning. We simply computed $I(Q)$ for an extended $Q$-range and rescaled it. We fitted $c_r$ and $c_F$ to minimize the mean squared error between simulated and experimental $F(Q)$.
\begin{equation}
    \mathcal{L}(c_r,c_F) = \frac{1}{N_Q}\sum_{i=1}^{N_Q}\left(c_F F_\mathrm{sim}(Q_i;c_r) - F_\mathrm{exp}(Q_i)\right)^2.
\end{equation}
When rescaling, the grid of $Q$-points shifts, so we interpolated linearly $F_\mathrm{sim}$ on the experimental grid. This worked well since the grid spacing was small, $\Delta Q = \SI{0.007}{\per\angstrom}$, and the lattice constant was only rescaled by a few percent. The errors we report are weighted profile residuals, i.e. intensity-normalized root mean squared errors:
\begin{equation}
    \label{eq:rwp}
    R_\mathrm{wp}[F(Q)] = \sqrt{\frac{\mathcal{L}(c_r,c_F)}{\frac{1}{N_Q}\sum_{i=1}^{N_Q}F_\mathrm{exp}^2(Q_i)}}.
\end{equation}

\subsection{Constructing Nanoparticles}
The candidate structures we compare against experimental data are Wulff constructions and spherical NPs. The Wulff constructions are made with the \texttt{WulffPack} Python software module \cite{rahm_wulffpack_2020}. The crystal symmetry is fcc and the surface energies are extracted from the Materials Project \cite{horton_accelerated_2025, jain_commentary_2013, tran_surface_2016}. The DFT settings used to calculate the surface energies are comparable to the settings used for labelling the MLIP dataset. To include also low Miller index Wulff constructions, we generate particles in three groups, including only facets where the maximum Miller index is $\max (h,k,l)=1,2,3$. For decahedra and icosahedra, we also include three different values of the \textit{twin energy} parameter in \texttt{WulffPack}: $0$, $0.25\gamma_{111}$, $0.5\gamma_{111}$, where $\gamma_{111}=0.710 \ \mathrm{J/m^2}$ is the surface energy of the $(h,k,l)=(1,1,1)$ facet -- the lowest energy facet. A non-zero twin energy allows for indentations of corners.
The spherical NPs are constructed by generating a large fcc cube, centering it on a single atom, and removing atoms farther from the center than the desired radius.
We include all unique truncations that result in a size between 50 and 50\! 000 atoms for both Wulff constructions and spheres. In total, we consider 3655 NPs: 2425 decahedral, 510 icosahedral, 342 octahedral, and 378 spherical NPs.
\subsection{Structure Sampling}
For idealized sampling, the atomic positions are used directly from the constructed nanoparticles to simulate scattering with the Debye equation.
For relaxed sampling, these idealized geometries are relaxed with the MLIP in the Atomic Simulation Environment (ASE)~\cite{hjorth_larsen_atomic_2017}. We use the L-BFGS optimizer and a \SI{0.01}{eV/\angstrom} maximum force tolerance. In the SI, we test different convergence tolerances and find this value to be optimal.
For MD sampling, we average the simulated $F(Q)$ from 10 structures, spaced \SI{1}{ps} apart, from the final part of a MLIP-driven MD trajectory. We perform the MD in ASE using the Langevin thermostat with a friction coefficient of \SI{10}{ps^{-1}}. For the vacuum-exposed(solvated) NPs, we use a timestep of \SI{5}{fs}(\SI{1}{fs}). We discard the initial \SI{20}{ps} as a temperature equilibration phase and use the final \SI{10}{ps} for structure sampling unless otherwise specified.
\subsection{Machine-Learned Interatomic Potential}
We employed an MLIP to accelerate geometry relaxations and molecular-dynamics sampling while retaining DFT-level accuracy. We fine-tuned the MACE-MATPES-PBE-0 foundation model~\cite{batatia_mace_2022, batatia_design_2022, kaplan_foundational_2025, batatia_foundation_2025} to reproduce DFT total energies and atomic forces from the curated dataset.

\subsubsection{Dataset}
The training dataset comprised diverse gold and water environments: bulk water, bulk gold (fcc, bcc and hcp), periodic slabs (all low Miller-index facets with $\max(h,k,l) \leq 3$ and strain between $-10\%$ and $+10\%$), and finite gold clusters (octahedral, decahedral and icosahedral) containing between 42 and 561 atoms. Systems that contained water were periodic and included only gold-water interfaces (no water-vacuum interfaces); water densities ranged from $0.8 \ \mathrm{g \ cm^{-3}}$ to $1.1 \ \mathrm{g \ cm^{-3}}$. Gold configurations were sampled at 300--700 K and water configurations at 300--500 K. Solvated clusters had approximately five water layers between periodic images, and slab systems had at least six to simulate bulk-like character. Slabs were at least 10 Å thick. 
In total, the dataset comprises 521 bulk gold and 330 bulk water systems along with 1080/195 gold slabs exposed to water/vacuum, and 764/501 NPs in vacuum/water.
The full dataset is available at~\cite{frost_bagger_vegge_2026}.

\subsubsection{Training}
The dataset was split 80/10/10 into training, validation, and test sets. We minimized a two-part mean squared error loss function with energy and force terms. Training proceeded in two stages with default settings: energy weights of 1 and 1000, a force weight of 100, and learning rates of $5 \times 10^{-4}$ followed by $1 \times 10^{-4}$. Training was stopped when the validation error ceased improving. Test-set root-mean-square errors were $0.7 \ \mathrm{meV}$ per atom for energies and $\SI{11.3}{meV\per\angstrom}$ for forces. Training and inference used software versions MACE 0.3.16 and PyTorch \cite{ansel_pytorch_2024} 2.12 with CUDA 13.2.

\subsubsection{DFT Settings}
Reference DFT calculations were performed using the Vienna Ab initio Simulation Package \cite{kresse_ab_1993, kresse_efficiency_1996, kresse_efficient_1996} (VASP, version 6.2.1) with the Perdew--Burke--Ernzerhof (PBE) exchange-correlation functional \cite{perdew_generalized_1996}. The valence electronic wave functions were expanded in a plane-wave basis with a kinetic-energy cutoff of 350 eV, and the projector-augmented-wave (PAW) method was used for the ionic cores \cite{blochl_projector_1994, kresse_ultrasoft_1999}. The electronic self-consistent iterations were converged to $10^{-6} \ \mathrm{eV}$. Brillouin-zone sampling was carried out on a $\Gamma$-centered mesh with a k-point density of $\SI{50}{\angstrom}$ along periodic directions; for non-periodic directions, corresponding to the slab-normal direction and all directions for NP systems, a single k-point was used. Calculations were performed with the precision flag in VASP set to PREC=Accurate.


\medskip
\textbf{Acknowledgements} \par 
The authors acknowledge support from the Pioneer Center for Accelerating P2X Materials Discovery (CAPeX), DNRF grant number P3, the Novo Nordisk Foundation Grant no. NNF24OC0089800, NNF22OC0078009, NNF23OC0081359 and NNF25OC0102677.
We acknowledge the MAX IV Laboratory for beamtime on the DanMAX beamline under 
proposal 20250365. Research conducted at MAX IV, a Swedish 
national user facility, is supported by Vetenskapsrådet (Swedish Research 
Council, VR) under contract 2018-07152, Vinnova (Swedish Governmental Agency for 
Innovation Systems) under contract 2018-04969 and Formas under contract 
2019-02496. DanMAX is funded by the NUFI grant no. 4059-00009B. The authors also acknowledge Aleksandra Smolska, Armin Asghari Alamdari, and Kristian Junker Andersen from the Department of Biological and Chemical Engineering, Aarhus University, Denmark, for their assistance with the experimental measurements.
%

\medskip

%
\bibliographystyle{MSP}








\bibliography{references}

@article{anker_autonomous_2026,
    title = {Autonomous {Synthesis} of {Nanoparticles} with {Target} {Scattering} {Patterns}},
    volume = {20},
    issn = {1936-0851},
    url = {https://doi.org/10.1021/acsnano.5c15488},
    doi = {10.1021/acsnano.5c15488},
    number = {8},
    urldate = {2026-05-27},
    journal = {ACS Nano},
    publisher = {American Chemical Society},
    author = {Anker, Andy S. and Jensen, Jonas H. and González-Duque, Miguel and Moreno, Rodrigo and Smolska, Aleksandra and Juelsholt, Mikkel and Hardion, Vincent and Jørgensen, Mads R. V. and Faíña, Andrés and Quinson, Jonathan and Støy, Kasper and Vegge, Tejs},
    month = mar,
    year = {2026},
    pages = {6767--6782},
}

@article{johansen_gpu-accelerated_2024,
    title = {A {GPU}-{Accelerated} {Open}-{Source} {Python} {Package} for {Calculating} {Powder} {Diffraction}, {Small}-{Angle}-, and {Total} {Scattering} with the {Debye} {Scattering} {Equation}},
    volume = {9},
    issn = {2475-9066},
    url = {https://joss.theoj.org/papers/10.21105/joss.06024},
    doi = {10.21105/joss.06024},
    language = {en},
    number = {94},
    urldate = {2026-06-28},
    journal = {Journal of Open Source Software},
    author = {Johansen, Frederik L. and Anker, Andy S. and Friis-Jensen, Ulrik and Dam, Erik B. and Jensen, Kirsten M. {\O}. and Selvan, Raghavendra},
    month = feb,
    year = {2024},
    pages = {6024},
}

@article{horton_accelerated_2025,
    title = {Accelerated data-driven materials science with the {Materials} {Project}},
    volume = {24},
    copyright = {2025 Springer Nature Limited},
    issn = {1476-4660},
    url = {https://www.nature.com/articles/s41563-025-02272-0},
    doi = {10.1038/s41563-025-02272-0},
    language = {en},
    number = {10},
    urldate = {2026-08-05},
    journal = {Nature Materials},
    publisher = {Nature Publishing Group},
    author = {Horton, Matthew K. and Huck, Patrick and Yang, Ruo Xi and Munro, Jason M. and Dwaraknath, Shyam and Ganose, Alex M. and Kingsbury, Ryan S. and Wen, Mingjian and Shen, Jimmy X. and Mathis, Tyler S. and Kaplan, Aaron D. and Berket, Karlo and Riebesell, Janosh and George, Janine and Rosen, Andrew S. and Spotte-Smith, Evan W. C. and McDermott, Matthew J. and Cohen, Orion A. and Dunn, Alex and Kuner, Matthew C. and Rignanese, Gian-Marco and Petretto, Guido and Waroquiers, David and Griffin, Sinead M. and Neaton, Jeffrey B. and Chrzan, Daryl C. and Asta, Mark and Hautier, Geoffroy and Cholia, Shreyas and Ceder, Gerbrand and Ong, Shyue Ping and Jain, Anubhav and Persson, Kristin A.},
    month = oct,
    year = {2025},
    pages = {1522--1532},
}

@article{jain_commentary_2013,
    title = {Commentary: {The} {Materials} {Project}: {A} materials genome approach to accelerating materials innovation},
    volume = {1},
    issn = {2166-532X},
    shorttitle = {Commentary},
    url = {https://doi.org/10.1063/1.4812323},
    doi = {10.1063/1.4812323},
    number = {1},
    urldate = {2026-08-05},
    journal = {APL Materials},
    author = {Jain, Anubhav and Ong, Shyue Ping and Hautier, Geoffroy and Chen, Wei and Richards, William Davidson and Dacek, Stephen and Cholia, Shreyas and Gunter, Dan and Skinner, David and Ceder, Gerbrand and Persson, Kristin A.},
    month = jul,
    year = {2013},
    pages = {011002},
}

@article{tran_surface_2016,
    title = {Surface energies of elemental crystals},
    volume = {3},
    issn = {2052-4463},
    url = {https://www.nature.com/articles/sdata201680},
    doi = {10.1038/sdata.2016.80},
    number = {1},
    urldate = {2024-10-19},
    journal = {Scientific Data 2016 3:1},
    publisher = {Nature Publishing Group},
    author = {Tran, Richard and Xu, Zihan and Radhakrishnan, Balachandran and Winston, Donald and Sun, Wenhao and Persson, Kristin A. and Ong, Shyue Ping},
    month = sep,
    year = {2016},
    pages = {1--13},
}

@article{rahm_wulffpack_2020,
    title = {{WulffPack}: {A} {Python} package for {Wulff} constructions},
    volume = {5},
    issn = {2475-9066},
    shorttitle = {{WulffPack}},
    url = {https://joss.theoj.org/papers/10.21105/joss.01944},
    doi = {10.21105/joss.01944},
    language = {en},
    number = {45},
    urldate = {2026-08-06},
    journal = {Journal of Open Source Software},
    author = {Rahm, J. Magnus and Erhart, Paul},
    month = jan,
    year = {2020},
    pages = {1944},
}

@article{kresse_ab_1993,
    title = {Ab initio molecular dynamics for liquid metals},
    volume = {47},
    url = {https://link.aps.org/doi/10.1103/PhysRevB.47.558},
    doi = {10.1103/PhysRevB.47.558},
    number = {1},
    urldate = {2026-08-15},
    journal = {Physical Review B},
    publisher = {American Physical Society},
    author = {Kresse, G. and Hafner, J.},
    month = jan,
    year = {1993},
    pages = {558--561},
}

@article{kresse_efficiency_1996,
    title = {Efficiency of ab-initio total energy calculations for metals and semiconductors using a plane-wave basis set},
    volume = {6},
    issn = {0927-0256},
    url = {https://www.sciencedirect.com/science/article/pii/0927025696000080},
    doi = {10.1016/0927-0256(96)00008-0},
    number = {1},
    urldate = {2026-08-15},
    journal = {Computational Materials Science},
    author = {Kresse, G. and Furthmüller, J.},
    month = jul,
    year = {1996},
    pages = {15--50},
}

@article{kresse_efficient_1996,
    title = {Efficient iterative schemes for ab initio total-energy calculations using a plane-wave basis set},
    volume = {54},
    url = {https://link.aps.org/doi/10.1103/PhysRevB.54.11169},
    doi = {10.1103/PhysRevB.54.11169},
    number = {16},
    urldate = {2026-08-15},
    journal = {Physical Review B},
    publisher = {American Physical Society},
    author = {Kresse, G. and Furthmüller, J.},
    month = oct,
    year = {1996},
    pages = {11169--11186},
}

@article{perdew_generalized_1996,
    title = {Generalized {Gradient} {Approximation} {Made} {Simple}},
    volume = {77},
    url = {https://link.aps.org/doi/10.1103/PhysRevLett.77.3865},
    doi = {10.1103/PhysRevLett.77.3865},
    number = {18},
    urldate = {2026-07-31},
    journal = {Physical Review Letters},
    publisher = {American Physical Society},
    author = {Perdew, John P. and Burke, Kieron and Ernzerhof, Matthias},
    month = oct,
    year = {1996},
    pages = {3865--3868},
}

@article{kresse_ultrasoft_1999,
    title = {From ultrasoft pseudopotentials to the projector augmented-wave method},
    volume = {59},
    url = {https://link.aps.org/doi/10.1103/PhysRevB.59.1758},
    doi = {10.1103/PhysRevB.59.1758},
    number = {3},
    urldate = {2026-08-15},
    journal = {Physical Review B},
    publisher = {American Physical Society},
    author = {Kresse, G. and Joubert, D.},
    month = jan,
    year = {1999},
    pages = {1758--1775},
}

@inproceedings{ansel_pytorch_2024,
    address = {New York, NY, USA},
    series = {{ASPLOS} '24},
    title = {{PyTorch} 2: {Faster} {Machine} {Learning} {Through} {Dynamic} {Python} {Bytecode} {Transformation} and {Graph} {Compilation}},
    volume = {2},
    isbn = {979-8-4007-0385-0},
    shorttitle = {{PyTorch} 2},
    url = {https://dl.acm.org/doi/10.1145/3620665.3640366},
    doi = {10.1145/3620665.3640366},
    urldate = {2026-08-15},
    booktitle = {Proceedings of the 29th {ACM} {International} {Conference} on {Architectural} {Support} for {Programming} {Languages} and {Operating} {Systems}, {Volume} 2},
    publisher = {Association for Computing Machinery},
    author = {Ansel, Jason and Yang, Edward and He, Horace and Gimelshein, Natalia and Jain, Animesh and Voznesensky, Michael and Bao, Bin and Bell, Peter and Berard, David and Burovski, Evgeni and Chauhan, Geeta and Chourdia, Anjali and Constable, Will and Desmaison, Alban and DeVito, Zachary and Ellison, Elias and Feng, Will and Gong, Jiong and Gschwind, Michael and Hirsh, Brian and Huang, Sherlock and Kalambarkar, Kshiteej and Kirsch, Laurent and Lazos, Michael and Lezcano, Mario and Liang, Yanbo and Liang, Jason and Lu, Yinghai and Luk, C. K. and Maher, Bert and Pan, Yunjie and Puhrsch, Christian and Reso, Matthias and Saroufim, Mark and Siraichi, Marcos Yukio and Suk, Helen and Zhang, Shunting and Suo, Michael and Tillet, Phil and Zhao, Xu and Wang, Eikan and Zhou, Keren and Zou, Richard and Wang, Xiaodong and Mathews, Ajit and Wen, William and Chanan, Gregory and Wu, Peng and Chintala, Soumith},
    month = apr,
    year = {2024},
    pages = {929--947},
}

@article{blochl_projector_1994,
    title = {Projector augmented-wave method},
    volume = {50},
    url = {https://link.aps.org/doi/10.1103/PhysRevB.50.17953},
    doi = {10.1103/PhysRevB.50.17953},
    number = {24},
    urldate = {2026-08-15},
    journal = {Physical Review B},
    publisher = {American Physical Society},
    author = {Blöchl, P. E.},
    month = dec,
    year = {1994},
    pages = {17953--17979},
}

@inproceedings{batatia_mace_2022,
    title = {{MACE}: {Higher} {Order} {Equivariant} {Message} {Passing} {Neural} {Networks} for {Fast} and {Accurate} {Force} {Fields}},
    volume = {35},
    booktitle = {Advances in {Neural} {Information} {Processing} {Systems}},
    publisher = {Curran Associates, Inc.},
    author = {Batatia, Ilyes and Kovacs, David P and Simm, Gregor and Ortner, Christoph and Csanyi, Gabor},
    editor = {Koyejo, S and Mohamed, S and Agarwal, A and Belgrave, D and Cho, K and Oh, A},
    year = {2022},
    pages = {11423--11436},
}

@misc{batatia_design_2022,
      title={The Design Space of E(3)-Equivariant Atom-Centered Interatomic Potentials}, 
      author={Ilyes Batatia and Simon Batzner and Dávid Péter Kovács and Albert Musaelian and Gregor N. C. Simm and Ralf Drautz and Christoph Ortner and Boris Kozinsky and Gábor Csányi},
      year={2022},
      eprint={2205.06643},
      archivePrefix={arXiv},
      primaryClass={stat.ML},
      url={https://arxiv.org/abs/2205.06643}, 
}

@misc{kaplan_foundational_2025,
    title = {A {Foundational} {Potential} {Energy} {Surface} {Dataset} for {Materials}},
    url = {http://arxiv.org/abs/2503.04070},
    doi = {10.48550/arXiv.2503.04070},
    urldate = {2026-08-15},
    publisher = {arXiv},
    author = {Kaplan, Aaron D. and Liu, Runze and Qi, Ji and Ko, Tsz Wai and Deng, Bowen and Riebesell, Janosh and Ceder, Gerbrand and Persson, Kristin A. and Ong, Shyue Ping},
    month = mar,
    year = {2025},
}

@article{batatia_foundation_2025,
    title = {A foundation model for atomistic materials chemistry},
    volume = {163},
    issn = {0021-9606},
    url = {https://doi.org/10.1063/5.0297006},
    doi = {10.1063/5.0297006},
    number = {18},
    urldate = {2026-08-15},
    journal = {The Journal of Chemical Physics},
    author = {Batatia, Ilyes and Benner, Philipp and Chiang, Yuan and Elena, Alin M. and Kovács, Dávid P. and Riebesell, Janosh and Advincula, Xavier R. and Asta, Mark and Avaylon, Matthew and Baldwin, William J. and Berger, Fabian and Bernstein, Noam and Bhowmik, Arghya and Bigi, Filippo and Blau, Samuel M. and Cărare, Vlad and Ceriotti, Michele and Chong, Sanggyu and Darby, James P. and De, Sandip and Della Pia, Flaviano and Deringer, Volker L. and Elijošius, Rokas and El-Machachi, Zakariya and Fako, Edvin and Falcioni, Fabio and Ferrari, Andrea C. and Gardner, John L. A. and Gawkowski, Mikołaj J. and Genreith-Schriever, Annalena and George, Janine and Goodall, Rhys E. A. and Grandel, Jonas and Grey, Clare P. and Grigorev, Petr and Han, Shuang and Handley, Will and Heenen, Hendrik H. and Hermansson, Kersti and Ho, Cheuk Hin and Hofmann, Stephan and Holm, Christian and Jaafar, Jad and Jakob, Konstantin S. and Jung, Hyunwook and Kapil, Venkat and Kaplan, Aaron D. and Karimitari, Nima and Kermode, James R. and Kourtis, Panagiotis and Kroupa, Namu and Kullgren, Jolla and Kuner, Matthew C. and Kuryla, Domantas and Liepuoniute, Guoda and Lin, Chen and Margraf, Johannes T. and Magdău, Ioan-Bogdan and Michaelides, Angelos and Moore, J. Harry and Naik, Aakash A. and Niblett, Samuel P. and Norwood, Sam Walton and O’Neill, Niamh and Ortner, Christoph and Persson, Kristin A. and Reuter, Karsten and Rosen, Andrew S. and Rosset, Louise A. M. and Schaaf, Lars L. and Schran, Christoph and Shi, Benjamin X. and Sivonxay, Eric and Stenczel, Tamás K. and Sutton, Christopher and Svahn, Viktor and Swinburne, Thomas D. and Tilly, Jules and van der Oord, Cas and Vargas, Santiago and Varga-Umbrich, Eszter and Vegge, Tejs and Vondrák, Martin and Wang, Yangshuai and Witt, William C. and Wolf, Thomas and Zills, Fabian and Csányi, Gábor},
    month = nov,
    year = {2025},
    pages = {184110},
}

@article{debye_zerstreuung_1915,
    title = {Zerstreuung von {Röntgenstrahlen}},
    volume = {351},
    issn = {1521-3889},
    url = {https://onlinelibrary.wiley.com/doi/abs/10.1002/andp.19153510606},
    doi = {10.1002/andp.19153510606},
    language = {en},
    number = {6},
    urldate = {2026-06-03},
    journal = {Annalen der Physik},
    author = {Debye, P.},
    year = {1915},
    pages = {809--823},
}

@article{debye_interferenz_1913,
    title = {Interferenz von {Röntgenstrahlen} und {Wärmebewegung}},
    volume = {348},
    copyright = {Copyright © 1914 WILEY-VCH Verlag GmbH \& Co. KGaA, Weinheim},
    issn = {1521-3889},
    url = {https://onlinelibrary.wiley.com/doi/abs/10.1002/andp.19133480105},
    doi = {10.1002/andp.19133480105},
    language = {en},
    number = {1},
    urldate = {2026-08-17},
    journal = {Annalen der Physik},
    author = {Debye, P.},
    year = {1913},
    pages = {49--92},
}

@article{waller_zur_1923,
    title = {Zur {Frage} der {Einwirkung} der {Wärmebewegung} auf die {Interferenz} von {Röntgenstrahlen}},
    volume = {17},
    issn = {0044-3328},
    url = {https://doi.org/10.1007/BF01328696},
    doi = {10.1007/BF01328696},
    language = {de},
    number = {1},
    urldate = {2026-08-17},
    journal = {Zeitschrift für Physik},
    author = {Waller, Ivar},
    month = dec,
    year = {1923},
    pages = {398--408},
}

@article{chupas_ra-pdf_2003,
    author = "Chupas, Peter J. and Qiu, Xiangyun and Hanson, Jonathan C. and Lee, Peter L. and Grey, Clare P. and Billinge, Simon J. L.",
    title = "{Rapid-acquisition pair distribution function (RA-PDF) analysis}",
    journal = "Journal of Applied Crystallography",
    year = "2003",
    volume = "36",
    number = "6",
    pages = "1342--1347",
    month = "Dec",
    doi = {10.1107/S0021889803017564},
    url = {https://doi.org/10.1107/S0021889803017564},
}

@article{kieffer_pyfai-calib_2020,
    author = "Kieffer, J. and Valls, V. and Blanc, N. and Hennig, C.",
    title = "{New tools for calibrating diffraction setups}",
    journal = "Journal of Synchrotron Radiation",
    year = "2020",
    volume = "27",
    number = "2",
    pages = "558--566",
    month = "Mar",
    doi = {10.1107/S1600577520000776},
    url = {https://doi.org/10.1107/S1600577520000776},
}

@article{burns_inclined-detector_2023,
    author = "Burns, Nicholas and Rahemtulla, Aly and Annett, Scott and Moreno, Beatriz and Kycia, Stefan",
    title = "{An inclined detector geometry for improved X-ray total scattering measurements}",
    journal = "Journal of Applied Crystallography",
    year = "2023",
    volume = "56",
    number = "2",
    pages = "510--518",
    month = "Apr",
    doi = {10.1107/S1600576723001747},
    url = {https://doi.org/10.1107/S1600576723001747},
}

@article{jensen_azint_2022,
    author = "Jensen, Alexander Bernthz and Christensen, Thorbj{\o}rn Erik K{\o}ppen and Weninger, Clemens and Birkedal, Henrik",
    title = "{Very large-scale diffraction investigations enabled by a matrix-multiplication facilitated radial and azimuthal integration algorithm: {\it MatFRAIA}}",
    journal = "Journal of Synchrotron Radiation",
    year = "2022",
    volume = "29",
    number = "6",
    pages = "1420--1428",
    month = "Nov",
    doi = {10.1107/S1600577522008232},
    url = {https://doi.org/10.1107/S1600577522008232},
}

@article{juhas_pdfgetx3_2013,
    author = "Juh{\'{a}}s, P. and Davis, T. and Farrow, C.~L. and Billinge, S.~J.~L.",
    title = "{{\it PDFgetX3}: a rapid and highly automatable program for processing powder diffraction data into total scattering pair distribution functions}",
    journal = "Journal of Applied Crystallography",
    year = "2013",
    volume = "46",
    number = "2",
    pages = "560--566",
    month = "Apr",
    doi = {10.1107/S0021889813005190},
    url = {https://doi.org/10.1107/S0021889813005190},
}

@article{juhas_diffpy-cmi_2015,
    author = "Juh{\'{a}}s, Pavol and Farrow, Christopher~L. and Yang, Xiaohao and Knox, Kevin~R. and Billinge, Simon~J.~L.",
    title = "{Complex modeling: a strategy and software program for combining multiple information sources to solve ill posed structure and nanostructure inverse problems}",
    journal = "Acta Crystallographica Section A",
    year = "2015",
    volume = "71",
    number = "6",
    pages = "562--568",
    month = "Nov",
    doi = {10.1107/S2053273315014473},
    url = {https://doi.org/10.1107/S2053273315014473},
}

@article{canty_science_2025,
    title = {Science acceleration and accessibility with self-driving labs},
    volume = {16},
    copyright = {2025 The Author(s)},
    issn = {2041-1723},
    url = {https://www.nature.com/articles/s41467-025-59231-1},
    doi = {10.1038/s41467-025-59231-1},
    language = {en},
    number = {1},
    urldate = {2026-08-31},
    journal = {Nature Communications},
    publisher = {Nature Publishing Group},
    author = {Canty, Richard B. and Bennett, Jeffrey A. and Brown, Keith A. and Buonassisi, Tonio and Kalinin, Sergei V. and Kitchin, John R. and Maruyama, Benji and Moore, Robert G. and Schrier, Joshua and Seifrid, Martin and Sun, Shijing and Vegge, Tejs and Abolhasani, Milad},
    month = apr,
    year = {2025},
    pages = {3856},
}

@article{stier_materials_2024,
    title = {Materials {Acceleration} {Platforms} ({MAPs}): {Accelerating} {Materials} {Research} and {Development} to {Meet} {Urgent} {Societal} {Challenges}},
    volume = {36},
    copyright = {© 2024 The Author(s). Advanced Materials published by Wiley-VCH GmbH},
    issn = {1521-4095},
    shorttitle = {Materials {Acceleration} {Platforms} ({MAPs})},
    url = {https://onlinelibrary.wiley.com/doi/abs/10.1002/adma.202407791},
    doi = {10.1002/adma.202407791},
    language = {en},
    number = {45},
    urldate = {2026-08-31},
    journal = {Advanced Materials},
    author = {Stier, Simon P. and Kreisbeck, Christoph and Ihssen, Holger and Popp, Matthias Albert and Hauch, Jens and Malek, Kourosh and Reynaud, Marine and Goumans, T.p.m. and Carlsson, Johan and Todorov, Ilian and Gold, Lukas and Räder, Andreas and Wenzel, Wolfgang and Bandesha, Shahbaz Tareq and Jacques, Philippe and Garcia-Moreno, Francisco and Arcelus, Oier and Friederich, Pascal and Clark, Simon and Maglione, Mario and Laukkanen, Anssi and Castelli, Ivano Eligio and Carrasco, Javier and Cabanas, Montserrat Casas and Stein, Helge Sören and Ozcan, Ozlem and Elbert, David and Reuter, Karsten and Scheurer, Christoph and Demura, Masahiko and Han, Sang Soo and Vegge, Tejs and Nakamae, Sawako and Fabrizio, Monica and Kozdras, Mark},
    year = {2024},
    pages = {2407791},
}

@article{vogler_autonomous_2024,
    title = {Autonomous {Battery} {Optimization} by {Deploying} {Distributed} {Experiments} and {Simulations}},
    volume = {14},
    issn = {1614-6840},
    url = {https://onlinelibrary.wiley.com/doi/abs/10.1002/aenm.202403263},
    doi = {10.1002/aenm.202403263},
    language = {en},
    number = {46},
    urldate = {2026-08-31},
    journal = {Advanced Energy Materials},
    author = {Vogler, Monika and Steensen, Simon Krarup and Ramírez, Francisco Fernando and Merker, Leon and Busk, Jonas and Carlsson, Johan Martin and Rieger, Laura Hannemose and Zhang, Bojing and Liot, François and Pizzi, Giovanni and Hanke, Felix and Flores, Eibar and Hajiyani, Hamidreza and Fuchs, Stefan and Sanin, Alexey and Gaberšček, Miran and Castelli, Ivano Eligio and Clark, Simon and Vegge, Tejs and Bhowmik, Arghya and Stein, Helge Sören},
    year = {2024},
    pages = {2403263},
}

@article{sjolin_perqueue_2024,
    title = {{PerQueue}: managing complex and dynamic workflows},
    volume = {3},
    issn = {2635-098X},
    shorttitle = {{PerQueue}},
    url = {https://doi.org/10.1039/d4dd00134f},
    doi = {10.1039/d4dd00134f},
    number = {9},
    urldate = {2026-08-31},
    journal = {Digital Discovery},
    author = {Sjølin, Benjamin Heckscher and Hansen, William Sandholt and Morin-Martinez, Armando Antonio and Petersen, Martin Hoffmann and Rieger, Laura Hannemose and Vegge, Tejs and García-Lastra, Juan Maria and Castelli, Ivano E.},
    month = sep,
    year = {2024},
    pages = {1832--1841},
}

@article{wulff_xxv_1901,
    title = {{XXV}. {Zur} {Frage} der {Geschwindigkeit} des {Wachsthums} und der {Auflösung} der {Krystallflächen}},
    volume = {34},
    copyright = {De Gruyter expressly reserves the right to use all content for commercial text and data mining within the meaning of Section 44b of the German Copyright Act.},
    issn = {2196-7105},
    url = {https://www.degruyterbrill.com/document/doi/10.1524/zkri.1901.34.1.449/html?lang=en},
    doi = {10.1524/zkri.1901.34.1.449},
    language = {de},
    number = {1-6},
    urldate = {2026-08-31},
    journal = {Zeitschrift für Kristallographie - Crystalline Materials},
    publisher = {De Gruyter},
    author = {Wulff, G.},
    month = dec,
    year = {1901},
    pages = {449--530},
}

@article{marks_modified_1983,
    title = {Modified {Wulff} constructions for twinned particles},
    volume = {61},
    issn = {0022-0248},
    doi = {10.1016/0022-0248(83)90184-7},
    number = {3},
    urldate = {2024-10-18},
    journal = {Journal of Crystal Growth},
    publisher = {North-Holland},
    author = {Marks, L. D.},
    month = apr,
    year = {1983},
    pages = {556--566},
}

@article{sedano_varo_gold_2024,
    title = {Gold {Nanoparticles} for {CO2} {Electroreduction}: {An} {Optimum} {Defined} by {Size} and {Shape}},
    volume = {146},
    issn = {0002-7863},
    url = {https://pubs.acs.org/doi/10.1021/jacs.3c10610},
    doi = {10.1021/jacs.3c10610},
    number = {3},
    journal = {Journal of the American Chemical Society},
    publisher = {American Chemical Society},
    author = {Sedano Varo, Esperanza and Egeberg Tankard, Rikke and Kryger-Baggesen, Joakim and Jinschek, Joerg and Helveg, Stig and Chorkendorff, Ib and Damsgaard, Christian Danvad and Kibsgaard, Jakob},
    month = jan,
    year = {2024},
    pages = {2015--2023},
}

@article{karlsen_importance_2026,
    title = {The importance of fit-for-purpose structure characterization in autonomous laboratories},
    copyright = {2026 Springer Nature Limited},
    issn = {2058-8437},
    url = {https://www.nature.com/articles/s41578-026-00953-z},
    doi = {10.1038/s41578-026-00953-z},
    language = {en},
    urldate = {2026-09-10},
    journal = {Nature Reviews Materials},
    publisher = {Nature Publishing Group},
    author = {Karlsen, Martin A. and Anker, Andy S.},
    month = aug,
    year = {2026},
    pages = {1--3},
}

@article{pozdnyakov_incompleteness_2020,
    title = {Incompleteness of {Atomic} {Structure} {Representations}},
    volume = {125},
    url = {https://link.aps.org/doi/10.1103/PhysRevLett.125.166001},
    doi = {10.1103/PhysRevLett.125.166001},
    number = {16},
    urldate = {2026-09-10},
    journal = {Physical Review Letters},
    publisher = {American Physical Society},
    author = {Pozdnyakov, Sergey N. and Willatt, Michael J. and Bartók, Albert P. and Ortner, Christoph and Csányi, Gábor and Ceriotti, Michele},
    month = oct,
    year = {2020},
    pages = {166001},
}

@article{maffettone_when_2025,
    title = {When can we trust structural models derived from pair distribution function measurements?},
    volume = {255},
    issn = {1359-6640},
    url = {https://doi.org/10.1039/d4fd00106k},
    doi = {10.1039/d4fd00106k},
    urldate = {2026-09-10},
    journal = {Faraday Discussions},
    author = {Maffettone, Phillip M. and Fletcher, William J. K. and Nicholas, Thomas C. and Deringer, Volker L. and Allison, Jane R. and Smith, Lorna J. and Goodwin, Andrew L.},
    month = jan,
    year = {2025},
    pages = {311--324},
}

@article{merchant_scaling_2023,
    title = {Scaling deep learning for materials discovery},
    volume = {624},
    copyright = {2023 The Author(s)},
    issn = {1476-4687},
    url = {https://www.nature.com/articles/s41586-023-06735-9},
    doi = {10.1038/s41586-023-06735-9},
    language = {en},
    number = {7990},
    urldate = {2026-09-10},
    journal = {Nature},
    publisher = {Nature Publishing Group},
    author = {Merchant, Amil and Batzner, Simon and Schoenholz, Samuel S. and Aykol, Muratahan and Cheon, Gowoon and Cubuk, Ekin Dogus},
    month = dec,
    year = {2023},
    pages = {80--85},
}

@article{szymanski_autonomous_2023,
    title = {An autonomous laboratory for the accelerated synthesis of inorganic materials},
    volume = {624},
    copyright = {2023 The Author(s)},
    issn = {1476-4687},
    url = {https://www.nature.com/articles/s41586-023-06734-w},
    doi = {10.1038/s41586-023-06734-w},
    language = {en},
    number = {7990},
    urldate = {2026-09-10},
    journal = {Nature},
    publisher = {Nature Publishing Group},
    author = {Szymanski, Nathan J. and Rendy, Bernardus and Fei, Yuxing and Kumar, Rishi E. and He, Tanjin and Milsted, David and McDermott, Matthew J. and Gallant, Max and Cubuk, Ekin Dogus and Merchant, Amil and Kim, Haegyeom and Jain, Anubhav and Bartel, Christopher J. and Persson, Kristin and Zeng, Yan and Ceder, Gerbrand},
    month = dec,
    year = {2023},
    pages = {86--91},
}

@article{leeman_challenges_2024,
    title = {Challenges in {High}-{Throughput} {Inorganic} {Materials} {Prediction} and {Autonomous} {Synthesis}},
    volume = {3},
    url = {https://link.aps.org/doi/10.1103/PRXEnergy.3.011002},
    doi = {10.1103/PRXEnergy.3.011002},
    number = {1},
    urldate = {2026-09-10},
    journal = {PRX Energy},
    publisher = {American Physical Society},
    author = {Leeman, Josh and Liu, Yuhan and Stiles, Joseph and Lee, Scott B. and Bhatt, Prajna and Schoop, Leslie M. and Palgrave, Robert G.},
    month = mar,
    year = {2024},
    pages = {011002},
}

@misc{frost_bagger_vegge_2026, title={Gold-water nanoparticle interface {DFT} dataset}, url={https://data.dtu.dk/articles/dataset/Gold-water_nanoparticle_interface_DFT_dataset/33543838/0}, DOI={10.11583/DTU.33543838}, abstractNote={<p dir="ltr">This dataset contains ab initio calculations of gold-water interfaces of extended surfaces and nanoparticles (NPs). It is meant for training machine learning interatomic potentials (MLIPs).</p><p dir="ltr">The dataset contains the total energy, interatomic forces, and stress tensor of 3391 atomic structures of gold and water evaluated with Density Functional Theory (DFT) at the generalized gradient approximation level (PBE). The dataset contains 521 bulk gold and 330 bulk water systems, 1080/195 gold slabs interfaced with water/vacuum, and 764/501 NPs surrounded by vacuum/water.</p><p dir="ltr">The AuNP1000 dataset (https://doi.org/10.11583/DTU.30480380) is a subset of this larger dataset but with the dispersion correction removed.</p><p dir="ltr">New to this dataset are</p><ul><li>Vacuum-exposed extended surfaces (slabs) of gold in a face centered cubic structure with low Miller index-facets: (110), (110), (111), (211). They are snapshots from MLIP-based molecular dynamics (MD) NVT trajectories with -10%,-5%,0%,+5%,+10% strain and a temperature of 300K, 500K, or 700K.</li><li>Water-exposed extended surfaces (slabs) of gold in a face centered cubic structure with low Miller index-facets: (110), (110), (111), (211). The column of water is approximately 20 Å tall with a cross-sectional area 50-100 Å^2, and a volume corresponding to a density of 1 g/cm^3 multiplied with 0.9, 1.0, 1.2 to sample different water densities. Surface strains of -10%,-5%,0%,+5%,+10% are sampled along with temperatures of 300 K, 500 K, 700 K from 100 ps MLIP-driven MD NVT trajectories.</li><li>Water-exposed gold nanoparticles/clusters with 55-309 atoms in octahedral, decahedral, and icosahedral geometries. Periodic images are separated by approximately 5 water layers. Sampled temperatures are 300-700K in MD NVT simulations with the water volume scaled by 0.9-1.2 from the equilibrium volume (estimated from short MD NPT simulation).</li><li>Bulk water (27 or 125 molecules) from MD NVT trajectories at temperatures of 300K, 350K,.,500K and a volume corresponding to a density of 1 g/cm^3 multiplied with 0.9,0.95,.,1.2.</li></ul><p dir="ltr">The units in the dataset file (gold-water-np-dataset.xyz) are eV, eV/Å, and eV/Å^3 for energies, interatomic forces, and stress tensor components, respectively.</p><p dir="ltr"><br></p><p dir="ltr"><b>DFT settings</b><br>The DFT was performed using the Vienna Ab initio Simulation Package (VASP) with the Perdew-Burke-Ernzerhof (PBE) exchange-correlation functional. A plane wave energy cutoff of 350 eV was used along with a k-point density of 50 Å in periodic directions (e.g. a mesh with (50Å/a, 50Å/b, 50Å/c) grid points for a fully periodic system with cell vector lengths a,b,c). The used INCAR is uploaded alongside the dataset.</p><p><br></p>}, publisher={Technical University of Denmark}, author={Frost, Emil Jermiin Pedersen and Bagger, Alexander and Vegge, Tejs}, year={2026}, month={Sep} }

@article{hjorth_larsen_atomic_2017,
    title = {The atomic simulation environment—a {Python} library for working with atoms},
    volume = {29},
    issn = {0953-8984},
    url = {https://doi.org/10.1088/1361-648X/aa680e},
    doi = {10.1088/1361-648X/aa680e},
    language = {en},
    number = {27},
    urldate = {2026-08-31},
    journal = {Journal of Physics: Condensed Matter},
    publisher = {IOP Publishing},
    author = {Hjorth Larsen, Ask and Jørgen Mortensen, Jens and Blomqvist, Jakob and Castelli, Ivano E and Christensen, Rune and Dułak, Marcin and Friis, Jesper and Groves, Michael N and Hammer, Bjørk and Hargus, Cory and Hermes, Eric D and Jennings, Paul C and Bjerre Jensen, Peter and Kermode, James and Kitchin, John R and Leonhard Kolsbjerg, Esben and Kubal, Joseph and Kaasbjerg, Kristen and Lysgaard, Steen and Bergmann Maronsson, Jón and Maxson, Tristan and Olsen, Thomas and Pastewka, Lars and Peterson, Andrew and Rostgaard, Carsten and Schiøtz, Jakob and Schütt, Ole and Strange, Mikkel and Thygesen, Kristian S and Vegge, Tejs and Vilhelmsen, Lasse and Walter, Michael and Zeng, Zhenhua and Jacobsen, Karsten W},
    month = jun,
    year = {2017},
    pages = {273002},
}

@misc{zarrouk_linear-scaling_2025,
    title = {Linear-scaling calculation of experimental observables for molecular augmented dynamics simulations},
    url = {http://arxiv.org/abs/2509.22388},
    doi = {10.48550/arXiv.2509.22388},
    urldate = {2026-09-17},
    publisher = {arXiv},
    author = {Zarrouk, Tigany and Caro, Miguel A.},
    month = sep,
    year = {2025},
    note = {arXiv:2509.22388 [cond-mat.mtrl-sci]},
}

@misc{zarrouk_molecular_2025,
    title = {Molecular augmented dynamics: {Generating} experimentally consistent atomistic structures by design},
    shorttitle = {Molecular augmented dynamics},
    url = {http://arxiv.org/abs/2508.17132},
    doi = {10.48550/arXiv.2508.17132},
    urldate = {2026-09-17},
    publisher = {arXiv},
    author = {Zarrouk, Tigany and Caro, Miguel A.},
    month = sep,
    year = {2025},
    note = {arXiv:2508.17132 [cond-mat.mtrl-sci]},
}

@misc{anker_autonomous_2025,
    title = {Autonomous interpretation of atomistic scattering data},
    url = {http://arxiv.org/abs/2510.05938},
    doi = {10.48550/arXiv.2510.05938},
    urldate = {2026-09-17},
    publisher = {arXiv},
    author = {Anker, Andy S. and Gardner, John L. A. and Rosset, Louise A. M. and Goodwin, Andrew L. and Deringer, Volker L.},
    month = oct,
    year = {2025},
    note = {arXiv:2510.05938 [cond-mat.mtrl-sci]},
}

\clearpage
\rhead{Supplemental Information}
\section*{Supplemental Information}
\appendix
\renewcommand\thefigure{S\arabic{figure}}
\renewcommand{\thetable}{S\arabic{table}}
\setcounter{figure}{0}
\setcounter{table}{0}
\section{Convergence of relaxed sampling}
To compute scattering from a minimum on the potential energy surface of our MACE MLIP, we first need to relax the geometry of the NPs. In practice, we set a convergence criterion, $F_\mathrm{max}$, based on the maximum norm within the atoms' force vectors.
\begin{equation}
    \max_i||\boldsymbol{F}_i|| = \max_i \sqrt{F_{i,x}^2+F_{i,y}^2+F_{i,z}^2} < F_\mathrm{max}.
\end{equation}
The geometry relaxation is considered converged once this inequality is met. The convergence criterion is chosen to balance accuracy of the scattering data and computational resources.

As $F_\mathrm{max}$ is decreased, the goodness-of-fit measure $R_\mathrm{wp}$ converges for NPs close to the minimum (see Figure~\ref{fig:relax_rwp_convergence}). This is important because it means the size estimates of experimental NPs will be consistent (see Figure~\ref{fig:relax_size_convergence}). Since $F_\mathrm{max}=\SI{0.01}{eV/\angstrom}$ and $F_\mathrm{max}=\SI{0.001}{eV/\angstrom}$ produce similar size estimates, we can save computational time without compromising accuracy. For a screening method in a funnel approach, consistent size estimates are the most important since they define the search range.

For NP geometry relaxation, we use the LBFGS method within ASE. The number of steps required for convergence depends on the morphology and size of the NP (see Figure~\ref{fig:relax_steps_hist} and \ref{fig:relax_steps_size}). For instance, spherical NPs require on average fewer steps than decahedral NPs. Convergence to $F_\mathrm{max}=\SI{0.001}{eV/\angstrom}$ requires about twice as many steps as $F_\mathrm{max}=\SI{0.01}{eV/\angstrom}$, meaning we can save about half the computation time without sacrificing size accuracy. Interestingly, we number of steps scales roughly with the cube root of the number of atoms, just like the NP diameter (see Figure~\ref{fig:relax_steps_size}).

\begin{figure}
    \centering
    \includegraphics[width=0.5\linewidth]{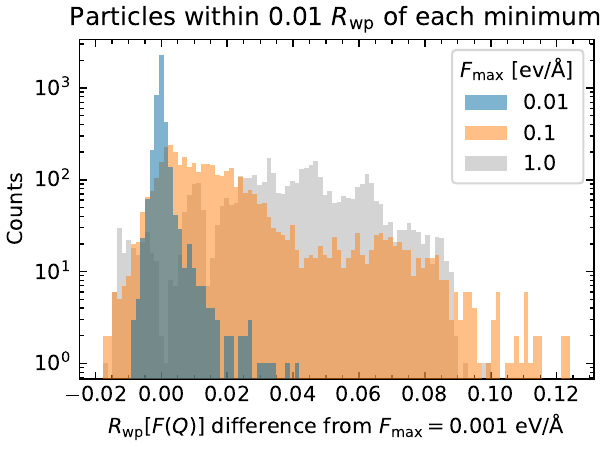}
    \caption{Histograms of differences in goodness-of-fit between NPs relaxed to $F_\mathrm{max}=1.0,0.1,\SI{0.01}{eV/\angstrom}$ and the reference: $F_\mathrm{max}=\SI{0.001}{eV/\angstrom}$. Included are only NPs within $\Delta R_\mathrm{wp} = 0.01$ of the best fitting NP in each experimental sample with any NP fit below $R_\mathrm{wp}=0.5$.}
    \label{fig:relax_rwp_convergence}
\end{figure}

\begin{figure}
    \centering
    \includegraphics[width=\linewidth]{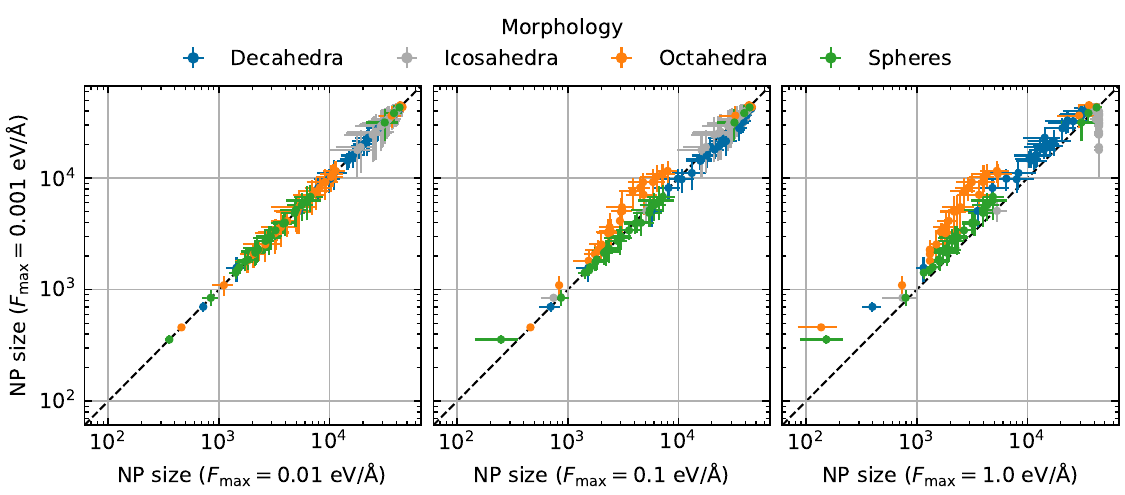}
    \caption{Parity plots between size estimates of all experimental samples with any NP fit below $R_\mathrm{wp}=0.5$ for different relaxation convergence criteria. The error bars signify the width of each minimum, i.e. the standard deviation of the distribution of sizes within $\Delta R_\mathrm{wp}=0.01$ of each minimum.
    }
    \label{fig:relax_size_convergence}
\end{figure}

\begin{figure}
    \centering
    \includegraphics[width=\linewidth]{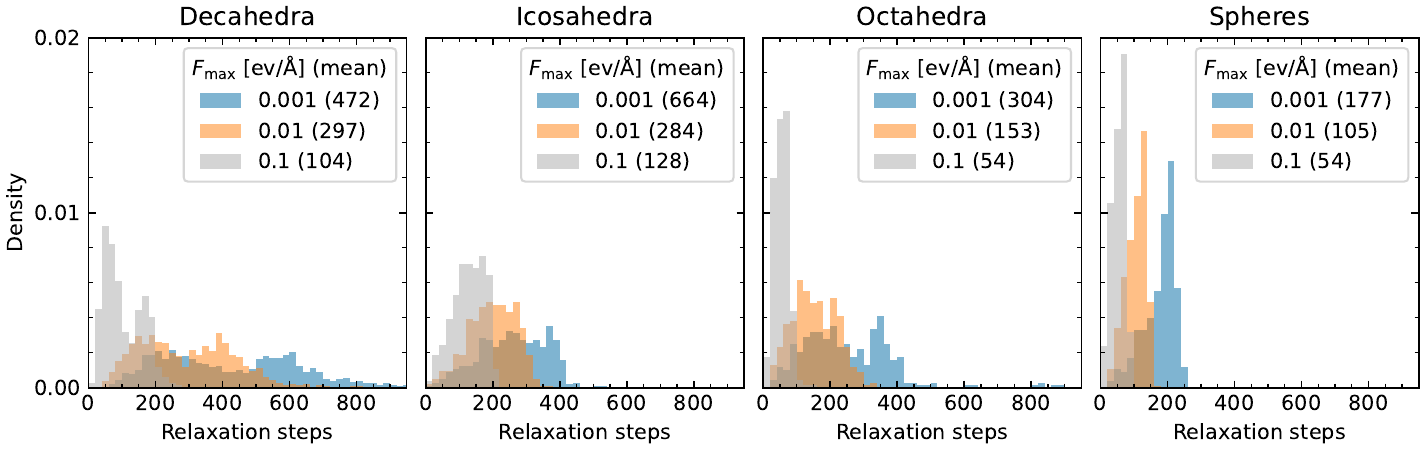}
    \caption{Distributions of the number of steps needed to relax NPs from different morphologies with L-BFGS at different maximum force thresholds with the MACE MLIP. The legend includes the average number of steps for each distribution. Some outliers are not visible in this size range (but included in the averages); they are visible in Figure \ref{fig:relax_steps_size}.}
    \label{fig:relax_steps_hist}
\end{figure}

\begin{figure}
    \centering
    \includegraphics[width=\linewidth]{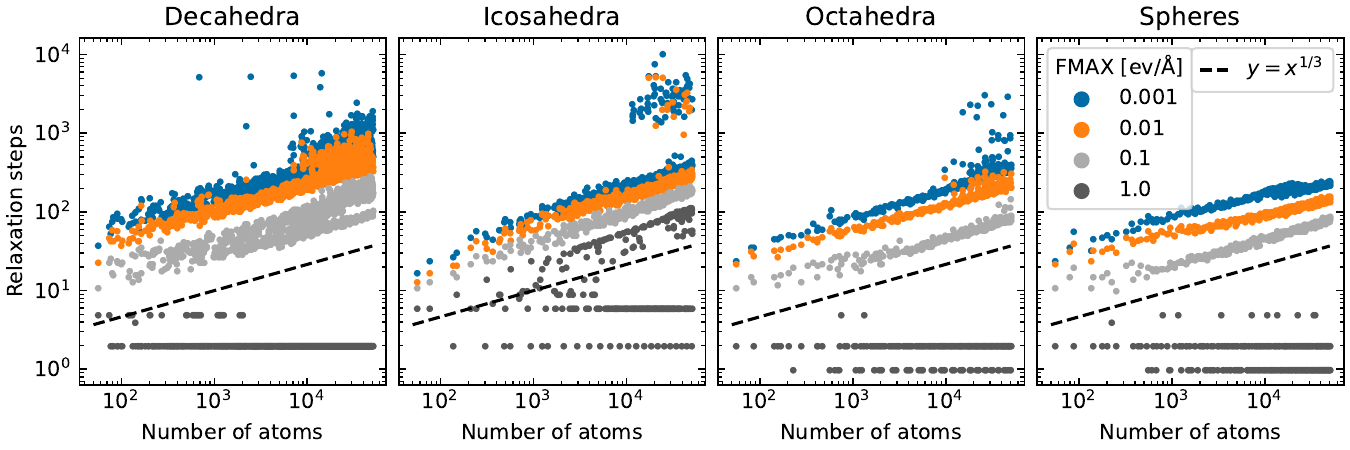}
    \caption{The number of steps needed to relax NPs from different morphologies with L-BFGS at different maximum force thresholds with the MACE MLIP. To guide the eye, the function $y=x^{1/3}$ that roughly matches the scaling, has been drawn as a dashed line. Hence, the number of steps scales approximately with the particle diameter.}
    \label{fig:relax_steps_size}
\end{figure}
\begin{figure}
    \centering
    \includegraphics[width=\linewidth]{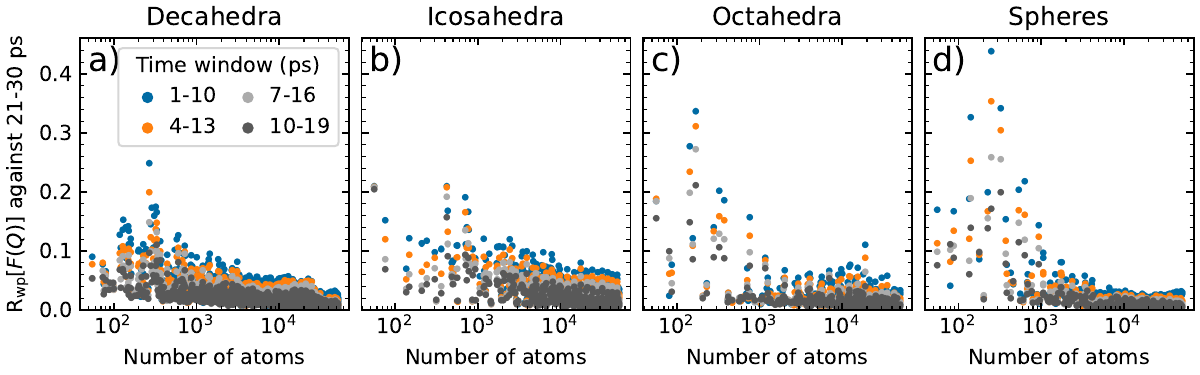}
    \caption{Convergence of $F(Q)$ average during MD, using $F(Q)$ averaged over the last 10 ps of each MD (21-30 ps time window) as references. The $F(Q)$ sampling frequency is 1 $\mathrm{ps^{-1}}$, giving 10 samples in each time window. The initial time window (1-10 ps) shows a large difference to the reference, while the 10-19 ps window is much closer. Small nanoparticles show a large spread, since they are heavily influenced by surface reconstruction. Large nanoparticles need longer time to equilibrate. The same data is reduced to violin plots in Figure~\ref{fig:fq_md_convergence_violin}.}
    \label{fig:fq_md_convergence}
\end{figure}

\begin{figure}
    \centering
    \includegraphics[width=\linewidth]{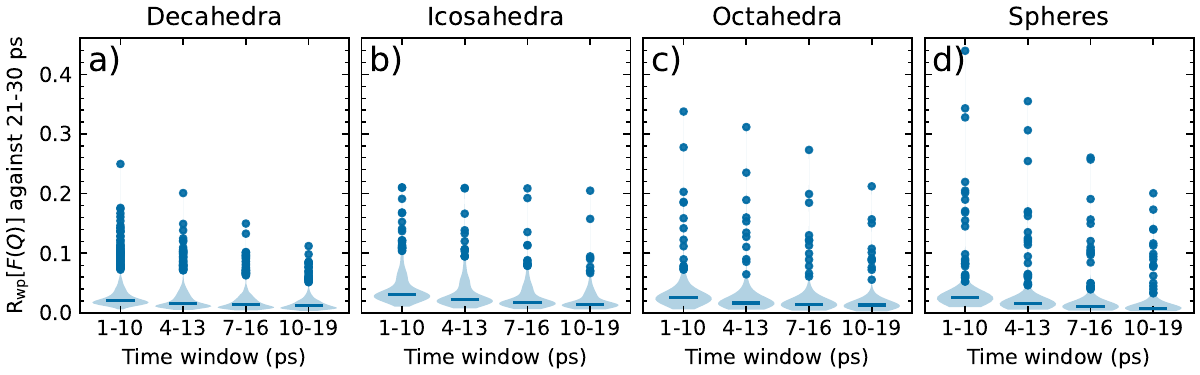}
    \caption{Convergence of $F(Q)$ average during MD, using $F(Q)$ averaged over the last 10 ps of each MD (21-30 ps time window) as references. The $F(Q)$ sampling frequency is 1 $\mathrm{ps^{-1}}$, giving 10 samples in each time window. Medians are marked with horizontal lines, and outliers that are more than three times the interquartile range above the 75th percentile are marked with points. The same data is size-resolved in Figure~\ref{fig:fq_md_convergence}, showing that the outliers are small nanoparticles.}
    \label{fig:fq_md_convergence_violin}
\end{figure}
\section{Temperature calibration of MD simulations in vacuum}
\label{sec:vacuum_water_calibration}
To sample the thermally averaged simulated scattering from solvated gold NPs, we found that performing the MD in vacuum at a reduced temperature was a good approximation. Instead of sampling with explicit water in the simulation cell and a temperature of $T=\SI{300}{K}$, we place the NP in vacuum at perform MD at $T=\SI{250}{K}$. Figure~\ref{fig:vacuum_temp_calibration} compares these two types of sampling for vacuum MD temperatures between \SI{100}{K} and \SI{300}{K}. Figure~\ref{fig:vacuum_temp_calibration}a shows a broad minimum in goodness-of-fit at a temperature slightly below $T=\SI{300}{K}$. For our simulations, we used a vacuum MD temperature of $T=\SI{250}{K}$; since the minimum is broad, the exact value is not crucial. Figure~\ref{fig:vacuum_temp_calibration}b shows that low/high temperature MD yields sharp/broad peaks. Water puts pressure on the gold atoms and dampens their motion, giving rise to sharper peaks. This looks similar to scattering from colder (more ordered) systems.

\begin{figure}
    \centering
    \includegraphics[width=\linewidth]{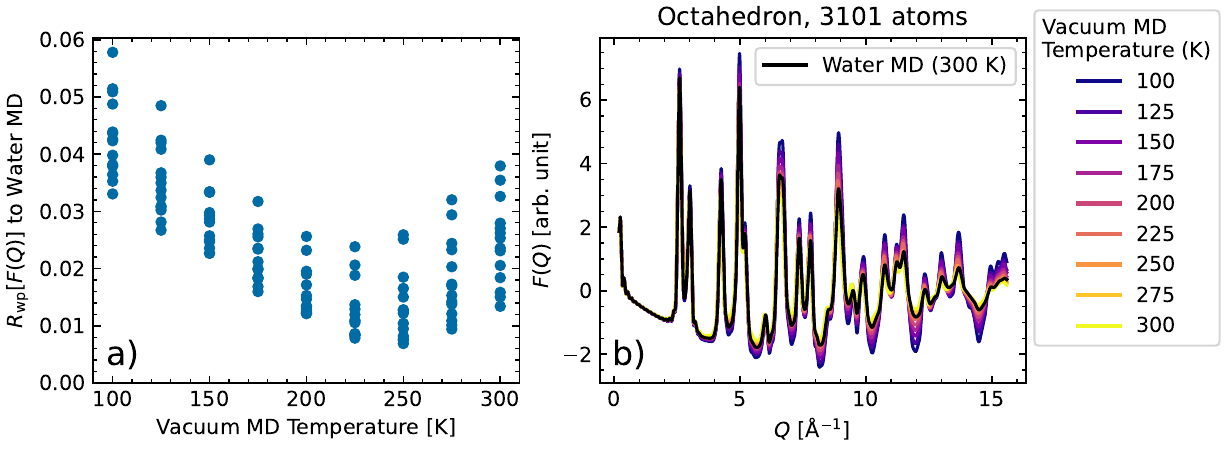}
    \caption{Comparison between thermally averaged simulated scattering from MD with/without explicit water in the simulation cell. The water MD thermostat is always set to $T=\SI{300}{K}$. (a) Goodness-of-fit between vacuum and water MD at different temperatures. The sampled particles are truncated octahedra with 201-3101 number of atoms. (b) An example of the difference in the reduced total scattering function, $F(Q)$, for a truncated octahedron with 3101 number of atoms. Peaks are sharper at low temperature and broader/smaller at higher temperature.}
    \label{fig:vacuum_temp_calibration}
\end{figure}
\begin{figure}
    \centering
    \includegraphics[width=0.8\linewidth]{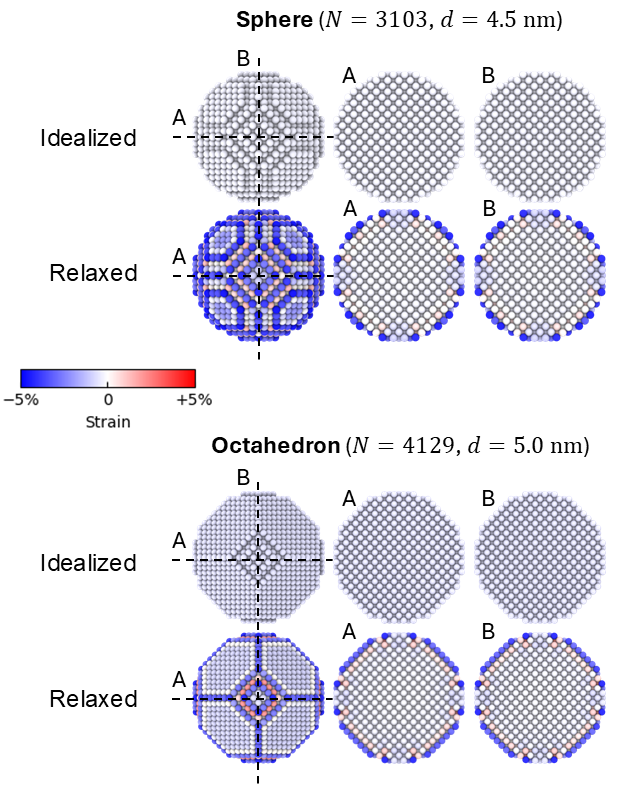}
    \caption{Cut-through views of the internal strain of the sphere and octahedron from Figure~\ref{fig:fq-size-scan}b. Note that the relative size is not correct. The idealized NP is scaled to minimize the mean squared displacement from the relaxed geometry. Dashed lines in the left column are the cutting planes used in the middle and right columns. Due to symmetry, the A and B views are identical. The number of atoms, \textit{N}, and the NP diameter, \textit{d}, are given in the panel titles. The local strain of atom \textit{i} is calculated as $\varepsilon_i = (\langle d_{ij}\rangle_j - d_\mathrm{bulk})/d_\mathrm{bulk}$ where the nearest-neighbor distance is averaged across all neighbors \textit{j} within \SI{3.5}{\angstrom} and $d_\mathrm{bulk}=\SI{2.938}{\angstrom}$ is calculated from a relaxed bulk fcc crystal with the MLIP.}
    \label{fig:strain_sphere_octa}
\end{figure}

\begin{figure}
    \centering
    \includegraphics[width=0.8\linewidth]{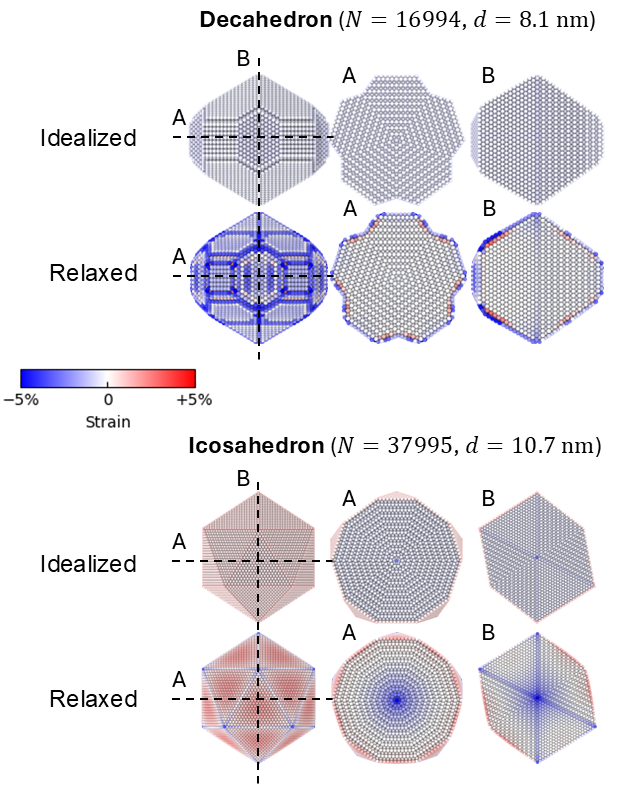}
    \caption{Cut-through views of the internal strain of the decahedron and icosahedron from Figure~\ref{fig:fq-size-scan}b. Note that the relative size is not correct. The idealized NP is scaled to minimize the mean squared displacement from the relaxed geometry. Dashed lines in the left column are the cutting planes used in the middle and right columns. The number of atoms, \textit{N}, and the NP diameter, \textit{d}, are given in the panel titles. The local strain of atom \textit{i} is calculated as $\varepsilon_i = (\langle d_{ij}\rangle_j - d_\mathrm{bulk})/d_\mathrm{bulk}$ where the nearest-neighbor distance is averaged across all neighbors \textit{j} within \SI{3.5}{\angstrom} and $d_\mathrm{bulk}=\SI{2.938}{\angstrom}$ is calculated from a relaxed bulk fcc crystal with the MLIP.}
    \label{fig:strain_deca_ico}
\end{figure}
\begin{figure}
    \centering
    \includegraphics[width=\linewidth]{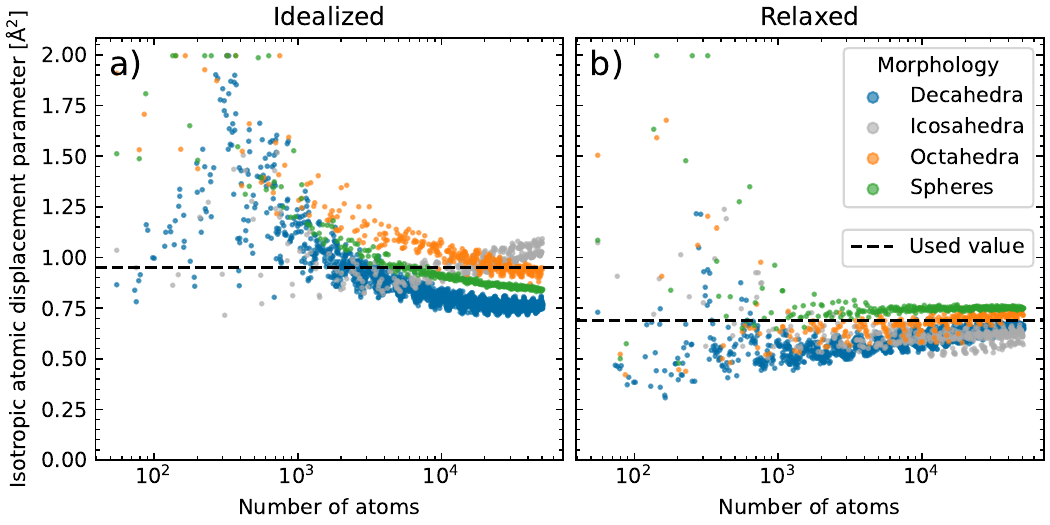}
    \caption{Best fit ADP values of a) idealized and b) relaxed NPs to the synthetic dataset of vacuum MD-sampled NPs. The dashed line indicates the ADP value used throughout this work; it is the minimum of Figure \ref{fig:rwp_biso_scan}.}
    \label{fig:fitted_biso}
\end{figure}

\begin{figure}
    \centering
    \includegraphics[width=0.6\linewidth]{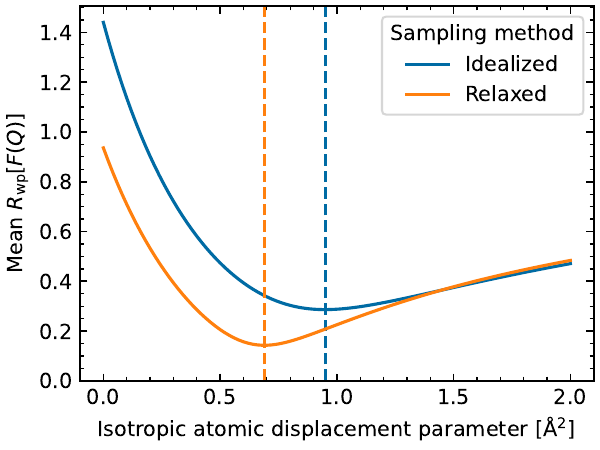}
    \caption{Average $R_\mathrm{wp}$ loss on the whole synthetic dataset of vacuum MD-sampled NPs for different ADP values. The ADP values at each minimum (marked with dashed lines) are used throughout this work, i.e. $B_\mathrm{iso}=\SI{0.95}{\angstrom^2}$ and $B_\mathrm{iso}=\SI{0.69}{\angstrom^2}$ for the idealized and relaxed sampling method, respectively.}
    \label{fig:rwp_biso_scan}
\end{figure}

\begin{figure}
    \centering
    \includegraphics[width=0.495\linewidth]{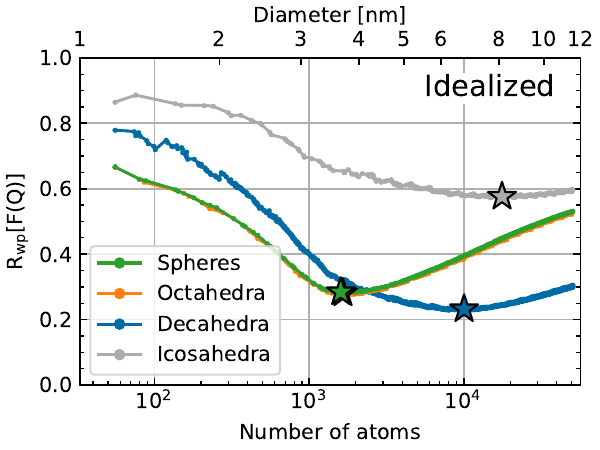}
    \includegraphics[width=0.495\linewidth]{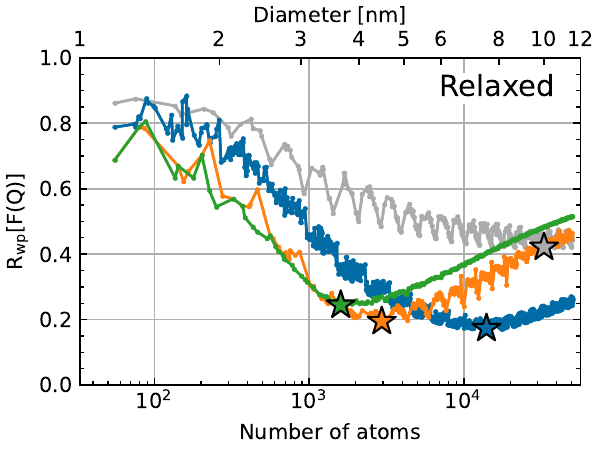}
    \caption{Idealized and relaxed sampled NPs with ADPs directly fitted to the experimental data used in Figure~\ref{fig:method-comparison}. These two plots should be directly compared to Figure~\ref{fig:method-comparison}a and \ref{fig:method-comparison}b. By fitting the ADP, size discrimination and morphology classification worsens; the octahedral and decahedral minima lie closer in goodness-of-fit than in Figure~\ref{fig:method-comparison}a and \ref{fig:method-comparison}b.}
    \label{fig:rwp_scan_free_biso}
\end{figure}

\begin{figure}
    \centering
    \includegraphics[width=\linewidth]{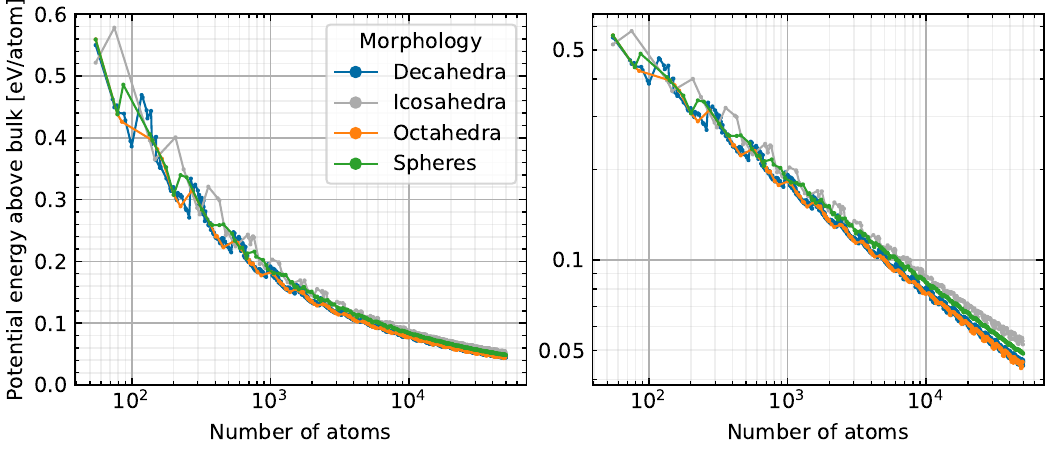}
    \caption{MACE potential energy of relaxed gold nanoparticles in vacuum referenced to the energy of bulk FCC. The left panel has a linear \textit{y}-axis, the right has a logarithmic. Only at the magic number sizes 55, 147, 309, and 561 atoms, are icosahedra the most stable of the four morphologies. Spherical NPs may only be stable when they coincide with octahedra. At large sizes, both icosahedra and spheres are unstable, while decahedral and octahedral NPs have similar energies.}
    \label{fig:energy_vs_size}
\end{figure}

\begin{figure}
    \centering
    \includegraphics[width=0.9\linewidth]{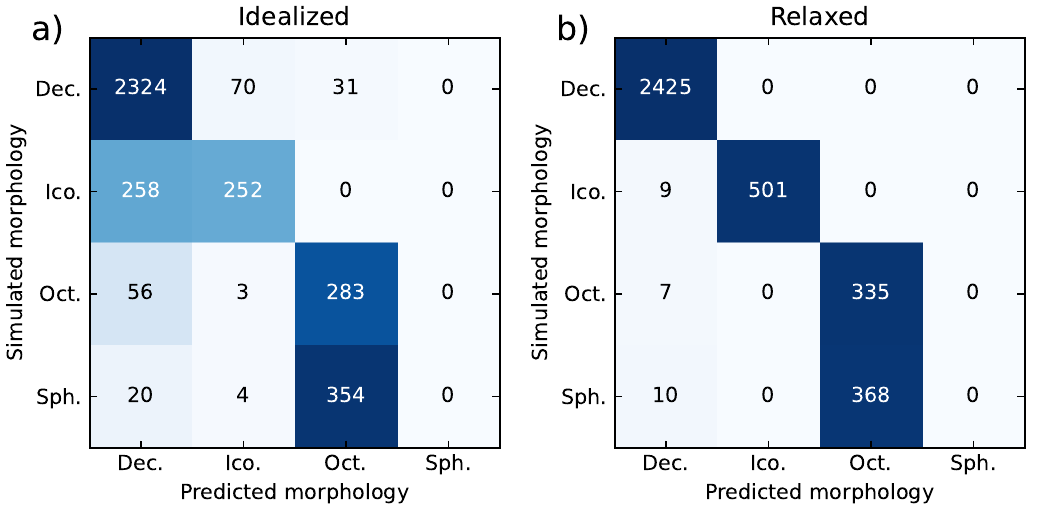}
    \caption{Confusion matrices for the morphology prediction of vacuum MD-sampled NPs using a) idealized and b) relaxed sampling. The idealized sampling struggles to separate decahedral NPs from the other morphologies. Neither approach predicts any NP to be spherical; they are mostly predicted to be octahedral, which only differs in surface termination.}
    \label{fig:confusion_matrix}
\end{figure}

\begin{figure}
    \centering
    \includegraphics[width=0.9\linewidth]{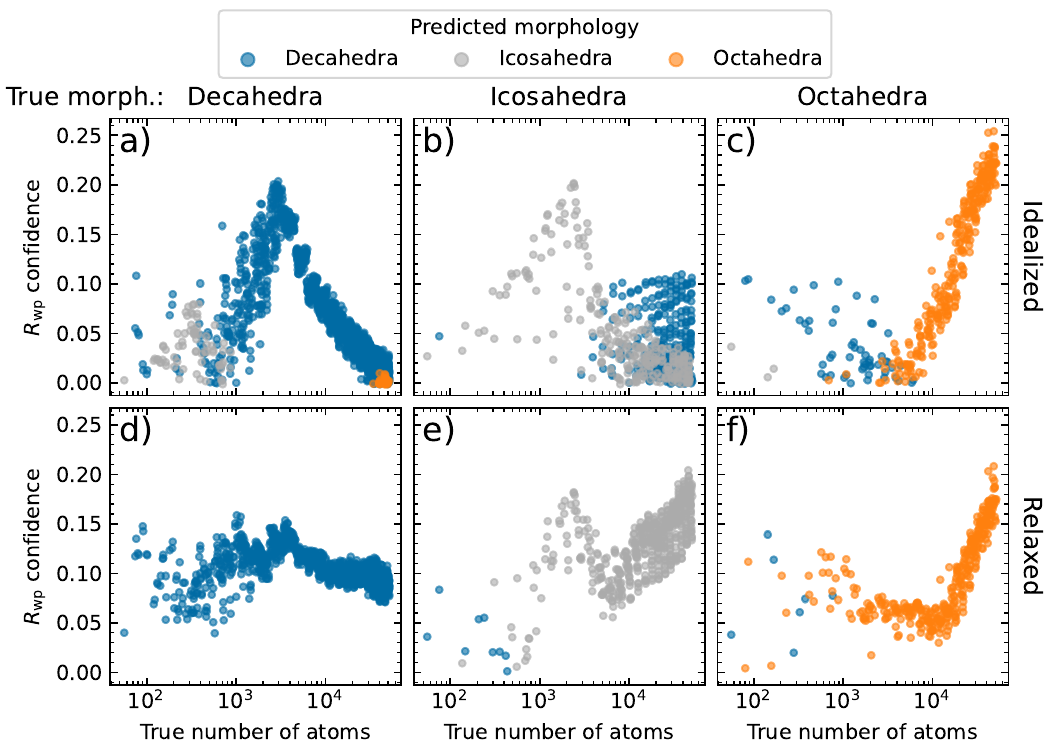}
    \caption{Morphology confidence as a function of particle size of vacuum MD-sampled NPs using a-c) idealized and d-f) relaxed sampling. Each column contains data for a single \textit{simulated} (true) morphology, while the colour shows the \textit{predicted} morphology. Morphology confidence is defined as the difference in $R_\mathrm{wp}[F(Q)]$ between the minima of the second best and best fitting morphology.}
    \label{fig:family_confidence}
\end{figure}

\begin{figure}
    \centering
    \includegraphics[width=\linewidth]{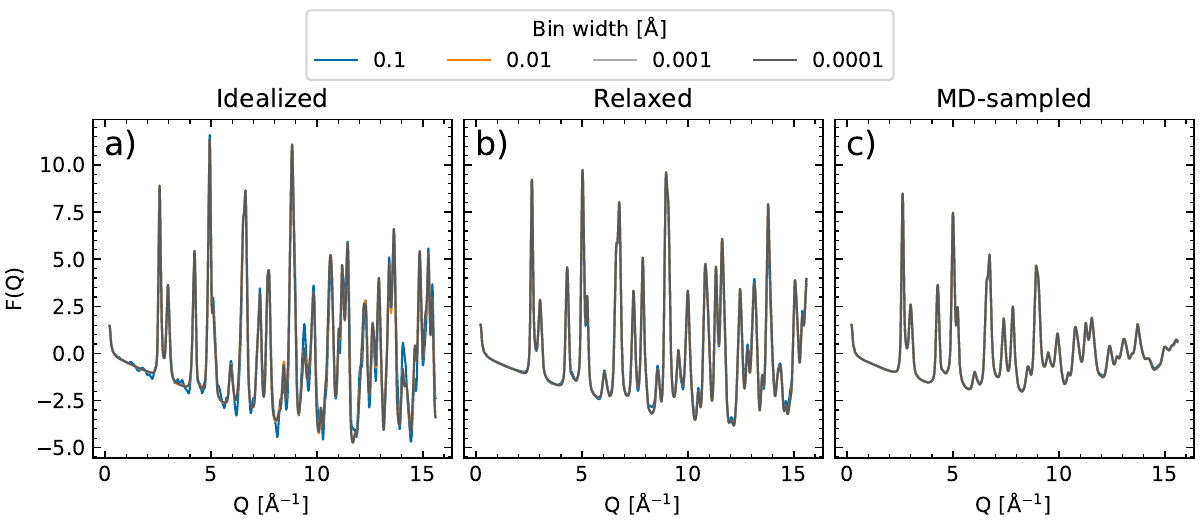}
    \caption{Convergence of the reduced total scattering function $F(Q)$ calculated from the Debye equation with binning of interatomic distances (Equation \eqref{eq:binned_debye_single}). The normalized mean squared errors from the unbinned $F(Q)$ are gathered in Table \ref{tab:binned_debye_convergence}. The nanoparticle structure is the decahedron shown in Figure \ref{fig:fq-size-scan}b, sampled with the three different methods: a) idealized, b) MACE relaxed, and c) MACE MD-sampled. The idealized nanoparticle shows the largest difference between bin widths due to very sharp peaks in the distance distribution; with MD-sampled nanoparticles, the distribution is much smoother.}
    \label{fig:binned_debye_convergence}
\end{figure}

\begin{table}
    \centering
    \begin{tabular}{lrrr}
        \toprule
         & \multicolumn{3}{c}{$\mathrm{R_{wp}}[F(Q)]$} \\
        \cmidrule(lr){2-4}
        Bin width [Å] & Idealized & Relaxed & MD-sampled \\
        \midrule
        0.1 & 0.193772 & 0.074879 & 0.025309 \\
        0.01 & 0.025109 & 0.000746 & 0.000340 \\
        0.001 & 0.001756 & 0.000047 & 0.000029 \\
        0.0001 & 0.000205 & 0.000004 & 0.000004 \\
        \bottomrule
    \end{tabular}
    \caption{Normalized mean squared error ($\mathrm{R_{wp}}$) between reduced total scattering functions $F(Q)$ calculated from the Debye equation with (Equation \eqref{eq:binned_debye_single}) and without (Equation \eqref{eq:debye_single}) binning of interatomic distances (see Figure \ref{fig:binned_debye_convergence}). As the bin width is reduced, the scattering function converges to the unbinned result.}
    \label{tab:binned_debye_convergence}
\end{table}

\begin{figure}
    \centering
    \includegraphics[width=\linewidth]{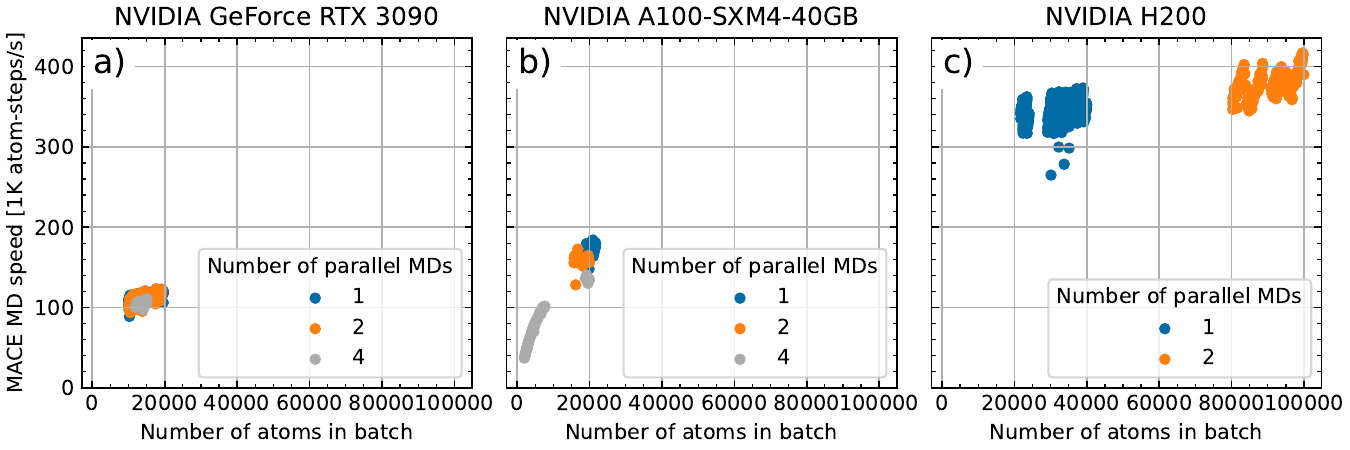}
    \caption{MACE molecular dynamics NVT timing on different GPUs (MACE v0.3.16, PyTorch v2.12.0 + CUDA v13.2). The model is a finetuned MACE-MATPES-PBE-0 foundation model. The MD is performed within ASE with the Langevin thermostat. For batches with few atoms, the throughput is reduced, partially due to the CPU-GPU communication overhead.}
    \label{fig:md_benchmark}
\end{figure}

\begin{figure}
    \centering
    \includegraphics[width=0.6\linewidth]{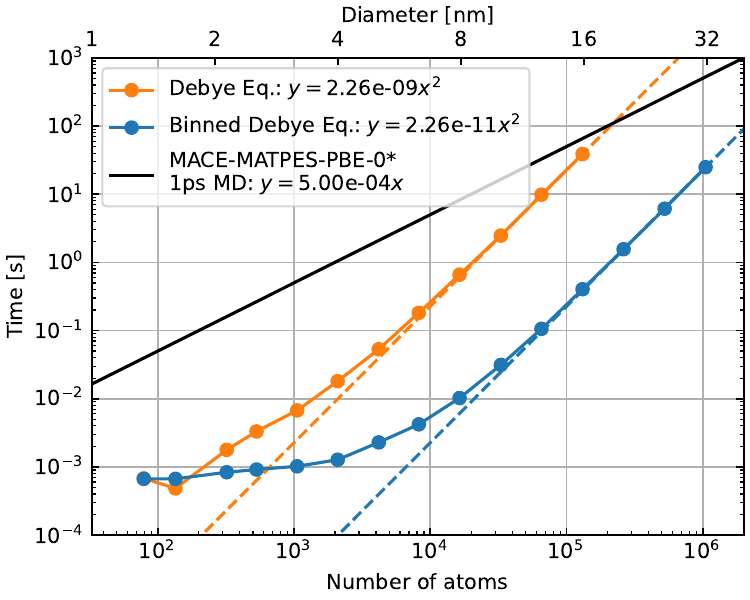}
    \caption{Benchmarking the Debye equation (Equation \eqref{eq:debye_single} and \eqref{eq:binned_debye_single}) on spherical NPs using an NVIDIA H200 GPU. The dashed lines are quadratic fits to the last five points on each of the two curves, demonstrating a $100\times$ speed-up at large sizes by binning distances. We used $N_Q=2067$ \textit{Q}-values, as in the experiments, and a bin width of \SI{0.001}{\angstrom}. For comparison, the time to perform \SI{1}{ps} MD with the MLIP (fine-tuned MACE-MATPES-PBE-0) is shown in black, assuming perfectly linear scaling with a throughput of \SI{400 000}{atoms \times steps / s} (see Figure~\ref{fig:md_benchmark}c) and \SI{5}{fs} time step, i.e. 200 energy/force evaluations.}
    \label{fig:debye_benchmark}
\end{figure}

\begin{figure}
    \centering
    \includegraphics[width=\linewidth]{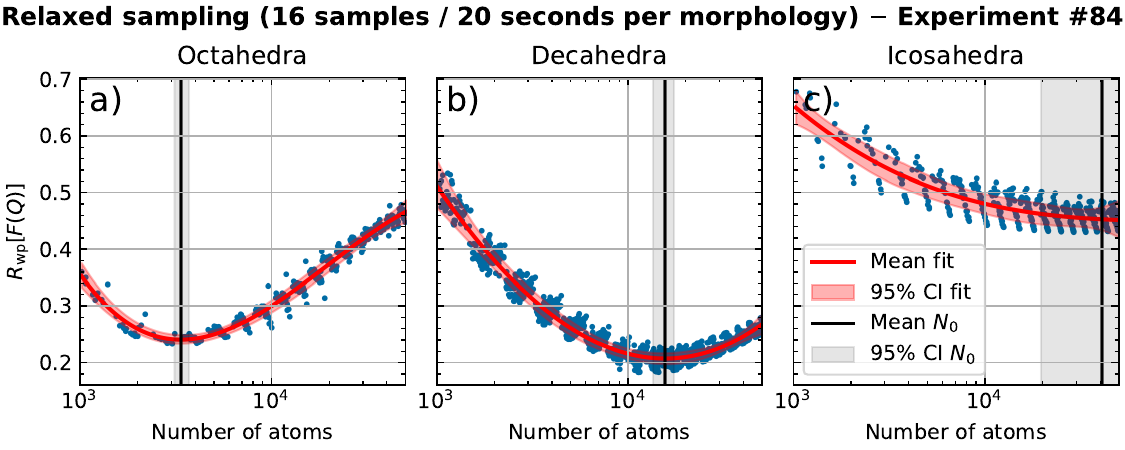}
    \caption{Estimating the optimal size of each morphology with relaxed sampling for the experimental sample considered in the main text. A subset of 16 randomly sampled structures from 16 logarithmically spaced bins is used to fit (regular least squares) a cubic polynomial: $R_\mathrm{wp}(t)=c_0+c_1t+c_2t^2+c_3t^3, \ t=\log N$. The time taken to relax the NPs with the finetuned MACE-MATPES-PBE-0 foundation model is 20 seconds for each morphology on 8 NVIDIA H200 GPUs. A bootstrapping procedure with 10\! 000 trials is used to determine the mean and 95\% confidence interval of the function value and the minimum position of the fit, $N_0$. The icosahedron minimum is poorly estimated since its flatter and lies at the edge of the range. Since the icosahedra are such a poor fit at the same time, we disregard these concerns.}
    \label{fig:relaxed_size_scan}
\end{figure}

\begin{figure}
    \centering
    \includegraphics[width=0.49\linewidth]{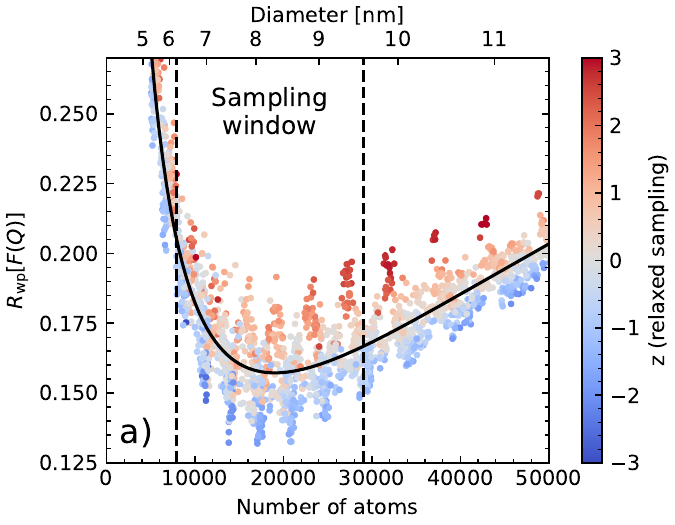}
    \hfill
    \includegraphics[width=0.49\linewidth]{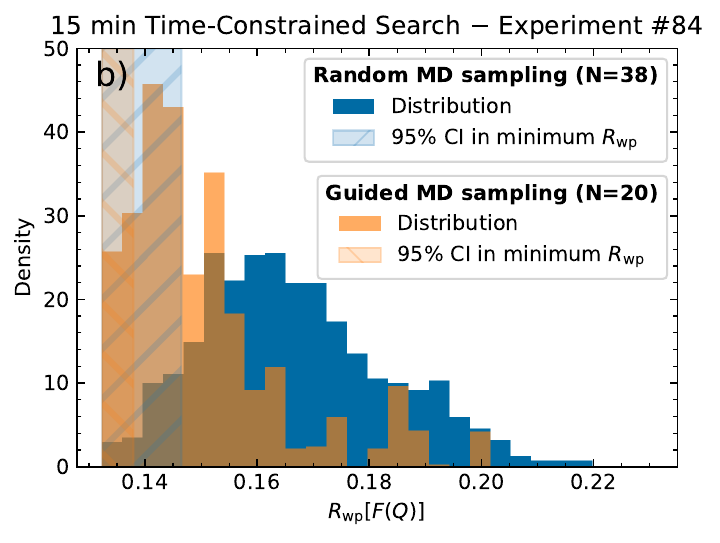}
    \caption{MD sampling of the decahedron minimum of the experimental sample considered in the main text. (a) Correlation of the scattering around the fit function between the relaxed NPs and the MD-sampled NPs. The points are coloured according to the normalized distance to the fit function of the \textit{relaxed} sampling in Figure~ \ref{fig:relaxed_size_scan}b (see Equation \eqref{eq:z}).
    For perfect correlation between relaxed and MD sampling, the colour would change uniformly top to bottom from red to blue which is almost the case. The sampling window used in (b) is marked with dashed lines and taken to be the interval in Figure~\ref{fig:relaxed_size_scan}b where $R_\mathrm{wp}(t)$ is within 0.02 of the minimum. (b) Comparison between random and guided sampling in a 15 minute time window on 8 NVIDIA H200 GPUs (including the $60 \, \mathrm{s}$ from the size estimation in Figure~\ref{fig:relaxed_size_scan}). Random sampling uses the time on 38 MD samples. Guided sampling uniformly samples candidates in the window, relaxes them, and forwards the samples with minimal $z$ to MD. In the figure, 10 times as many relaxed samples than MD samples are used, i.e. 20 MD samples and 200 relaxed samples. The bootstrapping is performed with 10\! 000 trials.}
    \label{fig:time_constrained_md}
\end{figure}

\end{document}